\documentclass[fleqn,usenatbib]{mnras}

\usepackage{newtxtext,newtxmath}
\usepackage[export]{adjustbox}
\usepackage{float}
\usepackage{hhline}
\usepackage{anyfontsize}
\usepackage{xcolor,soul}

\usepackage{indentfirst}
\usepackage{caption}
\usepackage{subcaption}
\usepackage[linesnumbered]{algorithm2e}
\usepackage{amsmath}	%
\usepackage{booktabs}
\usepackage[toc,page]{appendix}
\usepackage[backgroundcolor=yellow]{mdframed}

\usepackage[T1]{fontenc}
\usepackage{tabularx}
\usepackage[justification=centering]{caption}
\DeclareRobustCommand{\VAN}[3]{#2}
\let\VANthebibliography\thebibliography
\def\thebibliography{\DeclareRobustCommand{\VAN}[3]{##3}\VANthebibliography}

\usepackage{graphicx}	% Including figure files
\usepackage{mathtools}
\title[Halo Classification with NNs]{Radio Halo Detection in MWA Data using Deep Neural Networks and Generative Data Augmentation}

\author[A. K. Mishra et.al.]{\fontsize{14}{3}\selectfont
Ashutosh K. Mishra,$^{1,2}$\thanks{E-mail: ashutosh.mishra@ens.psl.eu (ENS)} Emma Tolley$^{2,3}$, Shreyam Parth Krishna$^{2,3}$, Jean-Paul Kneib$^{2}$\\
$^{1}$Department of Physics, École Normale Supérieure, PSL, Paris-75230, France\\
$^{2}$Institute of Physics, Laboratory of Astrophysics, École Polytechnique Fédérale de Lausanne (EPFL), Observatoire de Sauverny, Versoix, 1290, Switzerland\\
 $^{3}$SCITAS, École
 Polytechnique Fédérale de Lausanne (EPFL) , Lausanne, 1015, Switzerland}

\usepackage{physics}

\DeclareUnicodeCharacter{2212}{-}

\begin{document}
\label{firstpage}
\pagerange{\pageref{firstpage}--\pageref{lastpage}}
\maketitle

\begin{abstract}

Detecting diffuse radio emission, such as from halos, in galaxy clusters is crucial for understanding large-scale structure formation in the universe. Traditional methods, which rely on X-ray and Sunyaev-Zeldovich (SZ) cluster pre-selection, introduce biases that limit our understanding of the full population of diffuse radio sources. In this work, we provide a possible resolution for this astrophysical tension by developing a machine learning (ML) framework capable of unbiased detection of diffuse emission, using a limited real dataset like those from the Murchison Widefield Array (MWA). We generate for the first time radio halo images using Wasserstein Generative Adversarial Networks (WGANs) and Denoising Diffusion Probabilistic Models (DDPMs), and apply them to train a neural network classifier independent of pre-selection methods. The halo images generated by DDPMs are of higher quality than those produced by WGANs. The diffusion-supported classifier with a multi-head attention block achieved the best average validation accuracy of 95.93\% over 10 runs, using 36 clusters for training and 10 for testing, without further hyperparameter tuning. Using our classifier, we rediscovered 9/12 halos (75\% detection rate) from the MeerKAT Galaxy Cluster Legacy Survey (MGCLS) Catalogue, and 5/8 halos (63\% detection rate) from the Planck Sunyaev-Zeldovich Catalogue 2 (PSZ2) within the GaLactic and Extragalactic All-sky MWA (GLEAM) survey. In addition, we identify 11 potential new halos, minihalos, or candidates in the COSMOS field using XMM-chandra-detected clusters in GLEAM data. This work demonstrates the potential of ML for unbiased detection of diffuse emission and provides labeled datasets for further study. 

%Labeled data sets of radio halos are created for future use, and preliminary results indicate significant potential for application in other research problems where the availability of relevant samples is limited.
\end{abstract}


\maketitle
\section{Introduction}\label{intro}
Current evidence from the map of the observable universe suggests that the distribution of matter at large scales is not random, but forms a complex hierarchical structure called the Cosmic Web, composed of dark matter, gas, and galaxies  \cite[see][for a review]{2003_sander,2006ApJ...640..691V,Kravtsov_2012}. While galaxies group themselves in clusters and superclusters connected by the difficult-to-detect warm-hot intergalactic medium (WHIM) in the Cosmic Web filaments, the gas within these clusters known as the intracluster medium (ICM) emits observable X-rays \cite[e.g.][]{1976_mitchell,1977_Serlemitsos,1982_Forman,Adam_2017}. Studying these clusters of galaxies, their formation, and dynamics, holds the key to understanding the large-scale structure of the universe.

Magnetic fields on the order of $\mu$G level are found in these galaxy clusters \citep{Clarke_2001,Bonafede_2010} and are responsible for non-thermal radio synchrotron emission. Among these clusters, some exhibit large-scale ($\approx 1Mpc$) diffuse radio emission  \citep{2011_Van_Weeren)} in the form of centrally located radio \textit{halos} and peripherally located radio \textit{relics}. These diffuse sources inform us about the past and the evolution of these galaxy clusters by revealing details regarding the presence of large-scale magnetic fields and relativistic particles. In particular, their detection can give invaluable evidence about the merging state of the clusters or the so-called dynamical state of the clusters  \cite[e.g.][]{Cuciti_2015,Giancin_2017}.
However, the current methods for their detection are biased as the cluster samples selected via their Sunyaev-Zel'dovich \citep[SZ;][]{1972_SZ} signal show a higher diffuse emission detection rate than the X-ray-selected samples \cite[][]{Cuciti_2015,Kenda_2021}. This may be attributed to an X-ray selection bias towards relaxed cool-core clusters \citep{Andrade-Santos_2017}, or to different time-scales of SZ versus X-ray signal boosting due to mergers \cite[][]{Randall_2002,Poole_2007,Wik_2008}. This astrophysical tension drives the need for detection methods that are independent of these selection criteria.

The steep spectral indices\footnote{ The spectral index, $\alpha$ is defined as $S_{\nu} \propto \nu^{-\alpha}$ where $S_{\nu}$ is the flux
density at the frequency $\nu$.} ($\alpha \approx 1.3$) of these diffuse sources are largely due to the nature of injection spectra, and require low-frequency radio telescopes to ensure that the faint  diffuse radio emission is detected. Current radio interferometers such as the Giant Metrewave Radio Telescope \cite[GMRT;][]{GMRT1}, the Murchison Widefield Array \cite[MWA;][]{Tingay_2013}, the LOw-Frequency ARray  \cite[LOFAR;][]{van_2013}, the Australian Square Kilometre Array Pathfinder \cite[ASKAP;][]{PASA_2013}, and MeerKAT\footnote{Karoo Array Telescope} \citep{2016_Jonas} are starting to contribute to our understanding of these diffuse radio sources \citep{Wilber_2020}, simultaneously setting the stage for next-generation observations with the Square Kilometre Array Observatory (SKAO).

Halos are extended diffuse sources that generally follow the brightness distribution of the ICM and do not have optical counterparts \cite[see][for a review]{Van_2019, Paul_2023}. They come in two types: giant halos, which are megaparsec (Mpc)-sized structures typically found in massive disturbed galaxy clusters \cite[e.g.][]{Giovannini_1999,Cassano_2010}, and mini-halos, which are compact ($\approx 0.1-0.5 Mpc$) and located in the centers of relaxed cool-core clusters, particularly around the brightest cluster galaxies (BCGs) \cite[e.g.][]{Raja_2020, Biava_2024}. Their polarizations are mostly weak to be measured, mainly due to the beam depolarization effect of our instruments. The origin of halos is still debatable, but the most agreed-upon model is the merger-driven turbulent (re-) acceleration mechanism \citep{Brunetti_2016} for the seed electrons, which may have come into existence due to some hadronic processes \cite[][]{schlickeiser_1987, Ackermann_2014,Brunetti_2017}.

We adopt a simple classification of diffuse sources in halos and non-halos. For us, the class of halos includes even their candidates which are extended diffuse sources that are not yet confirmed but exhibit characteristics that suggest that they could be genuine halos. These characteristics typically include large-scale diffuse emission with no clear association to a compact radio source, central positioning within galaxy clusters, low surface brightness, and a lack of significant polarization \cite[e.g.][]{feretti_2012,brunetti_2014,Van_2019}. While further observational confirmation is often required to verify candidate halos, these traits form the basis for initial classification. To avoid speculating on the actual classification scheme and to concentrate on developing tools, this work focuses on detecting diffuse radio emission sources, particularly halos, and distinguishing them from other radio sources, such as Active Galactic Nuclei (AGNs). By automating this classification process, our approach can help the scientific community to accelerate the identification of central diffuse emission in clusters and facilitate follow-up observations on promising candidates.
To our knowledge, this is the first work done on radio halo classification. To achieve this, images produced by the MWA as part of the publicly released GaLactic and Extragalactic All-sky MWA (GLEAM) survey \cite[][]{Wayth_2015,2017_Hurley} have been utilized.

With an upsurge in the speed, size, and resolution of new radio observations from various telescopes, the automated classification of sources has become more important than ever, as the source density is getting higher because of the better sensitivity of the instruments. There have been many successes in the morphological classification of radio sources with the aid of machine learning (ML) using both Convolutional Neural Networks (CNNs)  \cite[see][]{Aniyan_2017,Alhassan_2018,Tang_2019,Samudre_2022,Masley_2021} and non-CNN approaches \citep{Lukic_2019,bowles2021e2,Scaife_2021,Sadeghi_2021,Wu_2019,Ntwae_2021} motivating the use of neural network (NN) for our work as well. More importantly, this introduces a new method for detecting diffuse emission that is blind to cluster selection biases, offering a potential resolution to the astrophysical tension previously discussed.

The automated detection of diffuse sources, particularly halos, faces a major challenge. The number of halos detected with the MWA is approximately one-third of those detected with the MGCLS despite the fact that we would expect synchrotron emission to be brighter at lower frequency. This is because MWA lacks longer baselines and operates at a lower observing frequency, resulting in the observation of contaminated and blended sources  \cite[for more information see][]{Stefan_2021}. This is precisely why we have chosen MWA data to train our machine learning models, making them robust enough to handle high-resolution datasets such as MGCLS. Traditional methods for addressing these issues in ML include image augmentation, such as rotation, reflection, pixel scaling, size scaling, Gaussian noise addition, and translation. They increase the number of images in the dataset to some extent. However, when the number of sources is low, on the order of 10, they require additional support from the ML model to work efficiently.

The solution lies in producing "realistic looking" or mock images of the halos and using them as additional samples to train and test the classifier. A similar technique has been used for the morphological classification of radio galaxies in \citet{rustige_2023}. Other studies have investigated the use of generative models to create images
of radio galaxies \citep{Ma_2018,Bastien_2021} using Variational AutoEncoders (VAE). GANs have been applied to astrophysical images by \citet{Schwainski_2017}.
For a semi-supervised GAN application to radio pulsars, see \citet{balakrishnan_2021}. In \citet{HACKSTEIN2023},various evaluation metrics are used to compare different generative models trained on optical galaxy images.

In this work, we develop an NN classifier to detect diffuse radio emission, focusing on radio halos. We explore generative data augmentation techniques with Generative Adversarial Networks (GANs) and Diffusion models to enhance classifier performance. We evaluate this classifier using real data from the MWA telescope, successfully rediscovering halos listed in the MGCLS Catalogue and the Planck Sunyaev-Zeldovich Catalogue 2 (PSZ2) within the GLEAM datasets. Additionally, we identify potential new halos, mini-halos, or candidates in the COSMOS field using XMM-Chandra-detected clusters. Our tool is designed for halo searches specifically in MWA data, using cataloged sources as references due to the data-dependent nature of ML. In future work, we aim to extend it for blind searches by incorporating multi-modal networks with data from radio, X-ray, optical, and other wavelengths.

This paper is structured as follows. In Section \ref{bkgd}, we review the two generative models, the standard metrics for image quality evaluation, and the multihead attention block to provide the necessary background. We introduce the data set used for training and validation in Section \ref{Dat}. In Section \ref{methods}, we outline the methods, including halo image generation for data augmentation and the classifier architectures, explaining their selection and relevance. We present the results and lead the discussion in Section \ref{ress} before we conclude in Section \ref{CON}. The details of normalization, network architectures, and hyperparameters are left in the appendix \ref{NetA}. The comparison results of the generative models are presented in the appendix \ref{GM_Res}.

$\Lambda$CDM Cosmology with $H_0 = 70$km s$^{-1}$ Mpc$^{-1}$, $\Omega_{\Lambda} = 0.7$, and $\Omega_m = 0.3$ is adopted throughout.
\section{Machine Learning Background}\label{bkgd}
In this paper, we explore two generative models, GANs and the Diffusion Model.  
 GANs perform one-shot generation and are fast, while diffusion models generate samples via many iterations, making them comparatively slower. Diffusion models are easier to control during training and are robust to small variations in the network architecture \citep{2022_Ruiz,debiased_diffusion_2023}. On the other hand, GANs optimize the adversarial min-max objective, which can lead to mode collapse. In contrast, the objective is simpler for diffusion models.
 
 We provide more details on our specific implementation of the generative models in the following sections. We briefly outline these generative models, standard metrics used for image quality evaluation, and the multi-head self-attention block used to enhance classifier performance to provide context.

\subsection{Denoising Diffusion Probabilistic Models} 
\begin{figure}
    \centering
    \includegraphics[width=\linewidth]{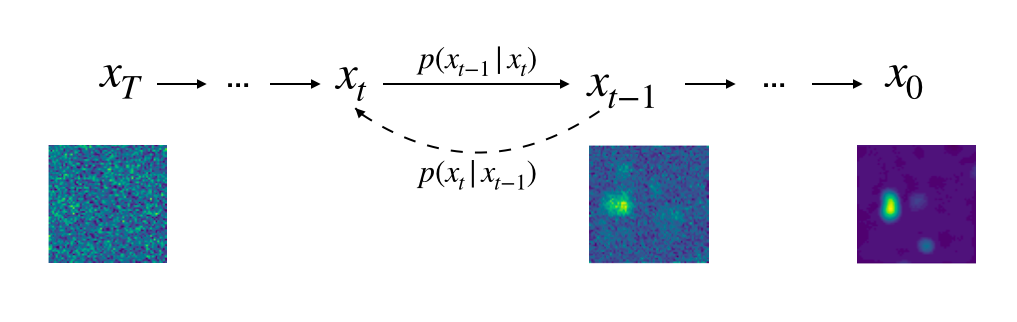}
    \caption{DDPM for Image Generation: 
    Starting from a highly noisy sample
    $x_T$, the model iteratively denoises the sample through intermediate states $x_t$ and $x_{t-1}$ using the reverse transition probability $p(x_{t-1}|x_{t})$, until it reconstructs the original data $x_0$. The dashed line indicates the forward process $p(x_t|x_{t-1})$, where noise is gradually added in the forward diffusion.}
    \label{fig:1}
\end{figure}

Diffusion models \cite[][]{diffusion}, as the name suggests, are probabilistic models motivated by ideas in nonequilibrium statistical mechanics. It is based on the principle of learning by iterative denoising. First, they perform forward diffusion where the random noise is gradually added to the data in diffusion steps of a Markov chain, and then reverse diffusion is learned to generate the required samples from the noise. This is schematically shown in Fig. (\ref{fig:1}).

The DDPM, as introduced by \citet{2020_Diff}, is a specific implementation of diffusion models. In DDPM, noisy versions of images are generated through a forward process defined by:
\begin{equation}
    \boldsymbol{x}_t = \sqrt{\overline{\alpha}_t} \boldsymbol{x}_0 + \sqrt{1 - \overline{\alpha}_t} \boldsymbol{\epsilon}
\end{equation}
with \(\alpha_t := 1 - \beta_t\) and \(\overline{\alpha}_t := \prod_{i=0}^{t} \alpha_i\). Here, \(\beta_t\) is a variance schedule parameter, which is linearly spaced between \(\beta_1 = 10^{-4}\) and \(\beta_T = 0.02\). The variable \(\boldsymbol{x}_0\) represents the original clean image, and \(\boldsymbol{x}_t\) represents the noisy image at time step \(t\), where \(t\) ranges from 0 to \(T\), with \(T\) being the total number of diffusion steps.

During the training stage, a neural network parameterized by \(\theta\) is used to predict the noise \(\boldsymbol{\epsilon}\) and is optimized using the following loss function:
\begin{equation}
    L(\theta) = \mathbb{E}_{t, \boldsymbol{x}, \boldsymbol{\epsilon}} \left[ \left\| \boldsymbol{\epsilon} - \boldsymbol{\epsilon}_\theta(\boldsymbol{x}_t, t) \right\|^2 \right]
\end{equation}
where \(\| \cdot \|\) denotes the \(L_2\) norm, and the expectation is over the denoising time-step \(t\) drawn from a uniform distribution. In this setup, \(T = 1000\). 

During the sampling stage, denoised versions of images are generated by reversing the diffusion process using the following formula:

\begin{equation}
    \boldsymbol{x}_{t-1} = \frac{1}{\sqrt{\alpha_t}} \left( \boldsymbol{x}_t - \frac{1 - \alpha_t}{\sqrt{1 - \overline{\alpha}_t}} \boldsymbol{\epsilon}_\theta(\boldsymbol{x}_t, t) \right) + \sigma_t \boldsymbol{z}
\end{equation}
where \(\sigma_t\) is a further variance schedule parameter and \(\boldsymbol{z}\) is standard Gaussian noise. This iterative denoising is guided by the neural network outputs, \(\boldsymbol{\epsilon}_\theta(\boldsymbol{x}_t, t)\), and the predefined schedule parameters. For this noise prediction, we use the U-Net Architecture as used in the original DDPM paper. We implement the architecture, hyperparameters, and training schedule parameters as described in the paper, with the sole modification being the number of groups in the GroupNormalization layer. The swish activation function \cite[][]{Ramchandran_2017} is used throughout. For more details, see \citet{2020_Diff}. 

\subsection{GANs}
 GANs \citep{goodfellow2014generative} consist of a pair of NNs coupled to each other in a competing Min-Max game. One network, called the \textit{generator G}, tries to create samples. The other, the \textit{discriminator D}, evaluates the generated samples to determine if they are real or fake. Once the system reaches an equilibrium \citep{Nash_1950,arora2017gans}, the network learns the true distribution, in principle. 

Let $p(\boldsymbol{x}|\theta)$ be the probability distribution function of radio flux density in halo images $\boldsymbol{x} \in [-1,1]^{H \times W}$(where H and W are the height and width of the images respectively), given the set of different halos parameterized by $\theta \in [0,9] \cap \mathbb{Z}$. The \textit{generator} gives samples from a distribution $p_G(\boldsymbol{x}|\theta$) which approaches the true distribution \textit{p} as the training progresses. Simultaneously, the \textit{discriminator} learns to differentiate between the samples from the generated distribution $p_G$ and the samples from the real distribution \textit{p}. In other words, the \textit{generator} and \textit{discriminator} are mappings $G : (\boldsymbol{z};\theta$) $\mapsto$ $G(\boldsymbol{z}|\theta$)  and $D :(\boldsymbol{x};\theta$) $\mapsto$ $D(\boldsymbol{x}|\theta$) $\in$ [0,1]. $\boldsymbol{z}$ represents the random noise which introduces stochasticity in the output generation making the network non-deterministic. It is usually drawn from either a uniform or normal distribution. While the input for \textit{the discriminator} is either a real sample $\boldsymbol{x}\sim p(\boldsymbol{x}|\theta$) or a fake sample i.e. sample generated by the \textit{generator}, $\boldsymbol{x}\sim p_G(\boldsymbol{x}|\theta$), and with these inputs, the \textit{discriminator} assigns a score to each or so to say a probability of the given input to be true or real input. One immediately notices in this description that this does not sound like a standard GAN, as the inputs are conditioned on a set of halos. It is a special GAN known as Conditional GAN \citep{mirza2014conditional}. The objective function is defined as 
\begin{align}
       (G', D') = \text{arg} \: \underset{G}{\text{min}}  \: \underset{D}{\text{max}} \: \mathcal{L}_{\text{GAN}} (G,D;\theta) 
\end{align}
where $\mathcal{L}_{\text{GAN}}$ is

\begin{align}
    \mathcal{L}_{\text{GAN}}(G,D;\theta) = \mathbb{E}_{\textbf{x} \sim p(\textbf{x})}[ \text{log} D(\textbf{x}|\theta)] \notag\\+ \mathbb{E}_{\textbf{z} \sim p_{\textbf{z}}(\textbf{z})}[\text{log}(1- D(G(\textbf{z}|\theta)|\theta)]
\end{align}
In this work, we use the Wasserstein GAN \citep[WGAN;][]{arjovsky2017wasserstein} which employs the Wasserstein distance (also called Earth-Mover's distance, which is continuous and differentiable everywhere) to solve the problem of mode collapse and unstable training, as it ends up providing useful gradients even when the generated $p_G$ and true distribution $p$ are far apart from each other.

To ensure that the training is stable, the \textit{discriminator} network's weights are constrained within a specific range in WGANs. To enforce this lipschitz constraint, \citet{gulrajani2017improved} advocate the use of the fact that a differentiable function f is 1-Lipschitz if its gradients are bounded by 1 everywhere implying the use of a penalty term and creating what we now call WGAN-GPs (WGANs with Gradient Penalty). They found that using gradient penalty in the \textit{discriminator's} loss term is a much better way to enforce the Lipschitz constraint, which also allows it to learn much more complex representations. The loss function of the \textit{discriminator} concerned is then given by
\begin{align}\label{Disc_loss}
    \mathcal{L}_{\text{WGAN-GP}}(G,D;\theta) = \mathbb{E}_{\textbf{x} \sim p(\textbf{x})}[ D(\textbf{x}|\theta)] \notag \\
    - \mathbb{E}_{\textbf{z} \sim p_{\textbf{z}}(\textbf{z})}[ D(G(\textbf{z}|\theta)|\theta)] +\notag\\ \lambda \mathbb{E}_{\tilde{\textbf{x}} \sim p_{\tilde{\textbf{x}}}} \left[ (\lVert{\nabla_{\tilde{\textbf{x}}} D(\tilde{\textbf{x}}|\theta)}\rVert_{2} - 1) ^{2}\right]
\end{align}
In Eq (\ref{Disc_loss}), $\tilde{\textbf{x}} = \alpha \textbf{x} + (1-\alpha)G(\textbf{z}|\theta)$ with a random number $\alpha \sim U[0,1]$ is an interpolated sample, and the factor $\lambda$ determines the strength of the gradient penalty. We use the value of the strength of gradient penalty $\lambda = 10$ in our work. It should be noted that the output of the \textit{discriminator} is no longer confined to $[0,1]$, it does not measure the probability of $\textbf{x} \sim p(\textbf{x}|\theta)$ but rather gives a number gauging how real the generated sample seems to it. In this sense, in WGANs, the \textit{discriminator} is called \textit{critic} whose goal is to maximize the difference between the scores (\textit{critic} scores) assigned by it for real and fake samples.

\begin{figure}
    \centering
    \includegraphics[width=\linewidth]{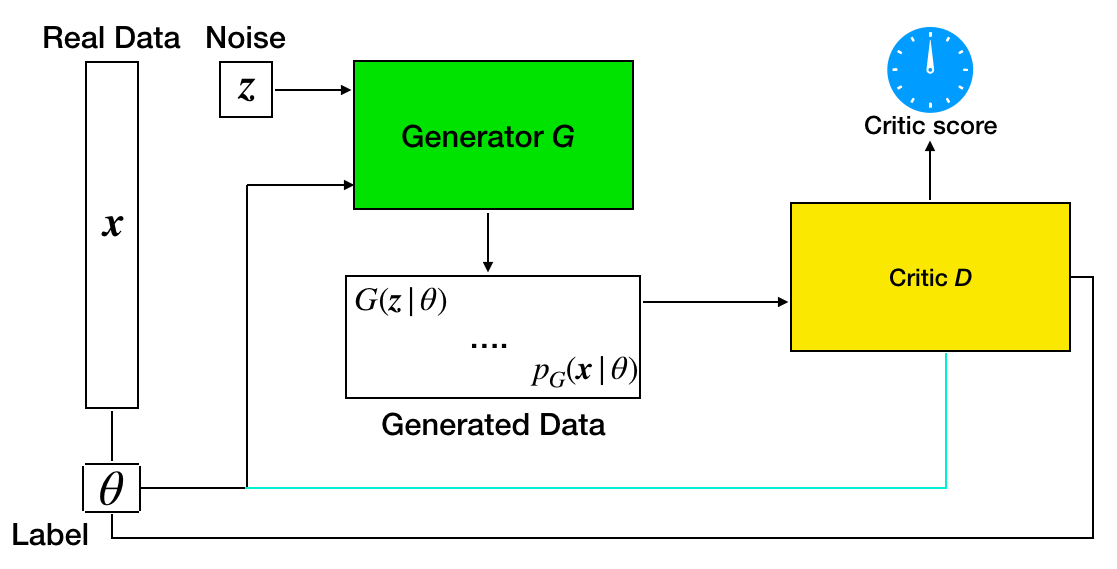}
    \caption{Schematic representation of WGAN used : The dots in the Generated Data box indicate that the generated image $G(\boldsymbol{z}|\theta)$ is a sample of the whole generated image distribution ($p_G(\boldsymbol{x}|\theta))$}
    \label{fig:2}
\end{figure}

Our architecture is inspired by DCGAN \cite[][]{radford2016unsupervised}. The workflow of our WGAN is schematically depicted in Fig. (\ref{fig:2}) and the details are listed in Appendix (\ref{NetA}).

\subsection{Metrics for Generated Image Quality}
One widely used metric in GAN research for evaluating generators is the Fréchet Inception Distance (FID) \cite[first introduced in][]{heusel2017gans}. The lower the FID score, the closer the generated images are to the real images in terms of their feature distributions, indicating higher image quality. The FID score is defined as:
\begin{equation}
\text{FID} = \|\mu_r - \mu_g\|_2^2 + \text{Tr}(\Sigma_r + \Sigma_g - 2(\Sigma_r \Sigma_g)^{\frac{1}{2}}) ,
\end{equation}
where \(\mu_r\) and \(\Sigma_r\) are the mean and covariance of the real data features, and \(\mu_g\) and \(\Sigma_g\) are the mean and covariance of the generated data features. We use InceptionV3 model \citep{szegedy2016rethinking} to extract the feature vectors for the images after resizing them to the input size (299,299,3) of the model. To avoid underestimation as suggested by \citet{heusel2017gans}, we use about 10000 samples to compute FID Score.

To remove the dependency of the metric on the sample size, Kernel Inception Distance (KID) was introduced in \citet{binkowski2018demystifying}. It is similar to FID but is based on the Maximum Mean Discrepancy (MMD) using features extracted from the Inception network. KID has several advantages, including unbiased estimation and better performance in some situations, especially when the number of samples is small. The MMD is defined as:
\begin{equation}
\text{MMD}^2 = \mathbb{E}_{x, x'} [k(x, x')] + \mathbb{E}_{y, y'} [k(y, y')] - 2 \mathbb{E}_{x, y} [k(x, y)],
\end{equation}
where \(k\) is a polynomial kernel, \(x\) and \(x'\) are features of real images, and \(y\) and \(y'\) are features of generated images. KID measures the difference between the means of these distributions, with lower values indicating higher quality of generated images. We use a sample size of 32 for KID calculations.
\subsection{Multi-Head Self Attention}\label{MHA}
\begin{figure}
    \centering
    \includegraphics[height = 5cm]{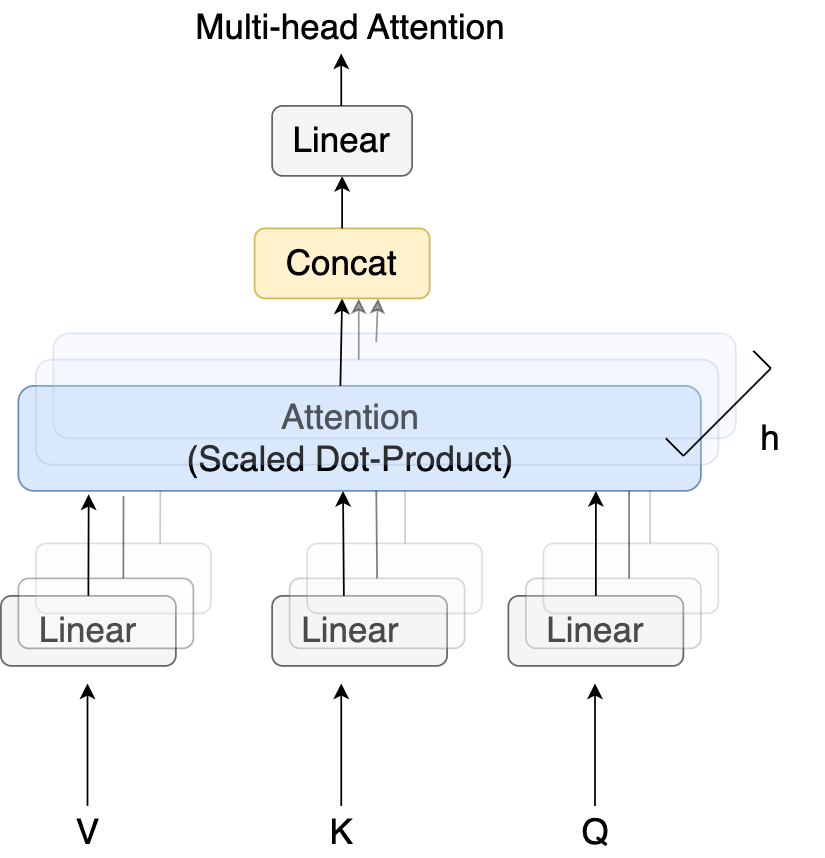}
    \caption{Multihead attention schematic, inspired from \citet{Vaswani_2017}: V, K, and Q are sets of values, keys, and queries respectively. The number of heads is denoted by h.} 
    \label{fig:app-MHA}
\end{figure}

Self-attention is a mechanism that computes representations of an input sequence by identifying and evaluating the relationships between various positions within the sequence. For each input vector $X$, a set of queries $Q$, keys $K$, and values $V$ are obtained by
$$Q = W_q X, \qquad V= W_v X , \qquad  K = W_k X.$$
In this context, $W_q, W_k,$ and $ W_v$ are weight matrices that linearly transform the input vector X into queries, keys, and values, respectively.  Here, "self-attention" implies that each element in the sequence is represented by the same input vector 
X, allowing the model to assess and incorporate information from different positions within the sequence. \citet{Vaswani_2017} defines attention as the following:
\begin{equation}\label{Eq1}
    \text{Attention}(Q,K,V) = \text{softmax}\bigr(QK^T /\sqrt{d_k} \bigl)V
\end{equation}
where the input consists of queries and keys of dimension $d_k$, and values of dimension $d_v$ (T represents the transpose operation here).
The same paper also introduced multi-head attention where one defines $h$ representations of ($Q$, $K$, $V$), computes self-attention for each of them, and concatenates them to get the final result. 
In the original work, linear or dense layers are used to calculate the h representations. However, these representations can also be calculated with convolutional layers \citep{Bairouk_2023}. 
In this work, we limit ourselves to using linear projections for classification, although convolutional projections can be beneficial for handling more complex datasets. Furthermore, we use $h = 8$ or $h = 4$ parallel attention layers (heads). A schematic for the architecture of the multihead attention is shown in Fig. (\ref{fig:app-MHA}). More details on its importance and architectural use in classification networks are discussed in section \ref{ress}.

\section{Data}\label{Dat}
\begin{figure}
\centering
\begin{tabular}{cc}
Halo & Candidate Halo \\
\includegraphics[width=4cm]{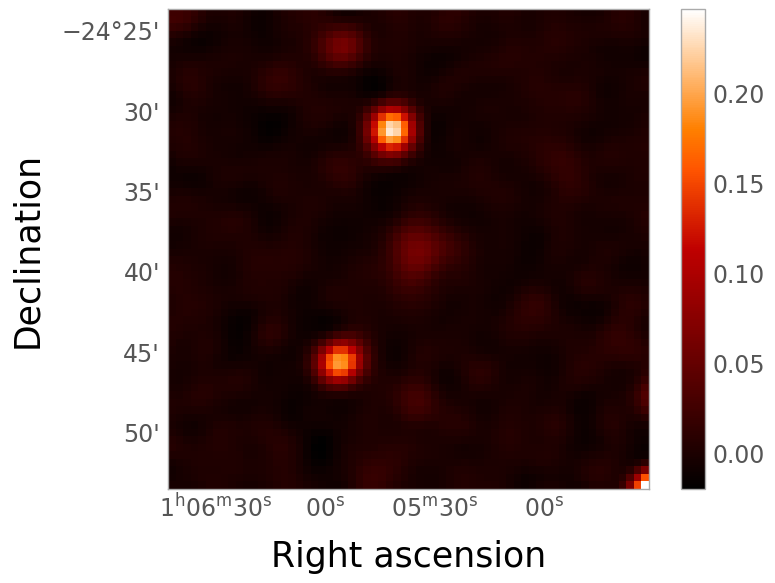} & \includegraphics[width=4cm]{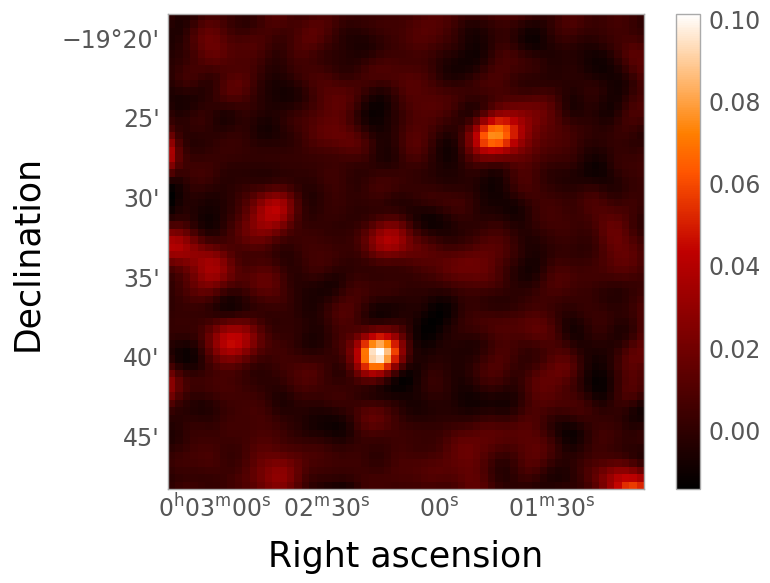}\\
Minihalo & Candidate Minihalo \\
\includegraphics[width=4cm]{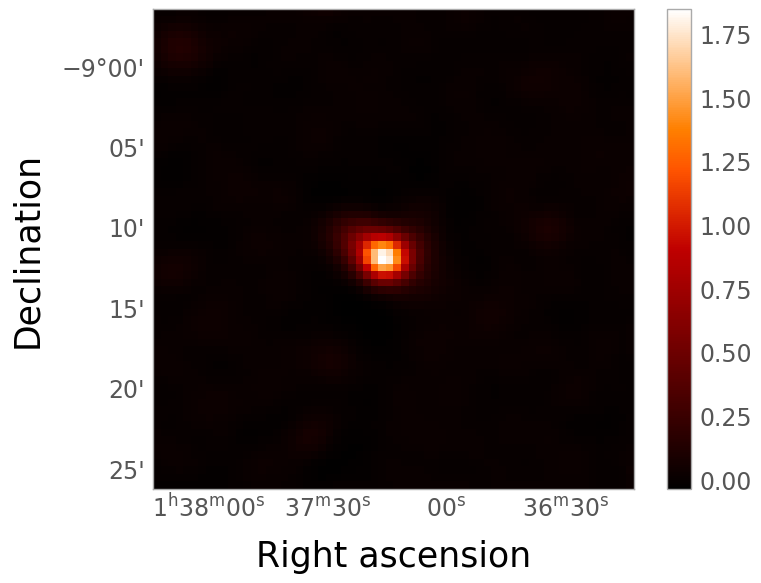} &
\includegraphics[width=4cm]{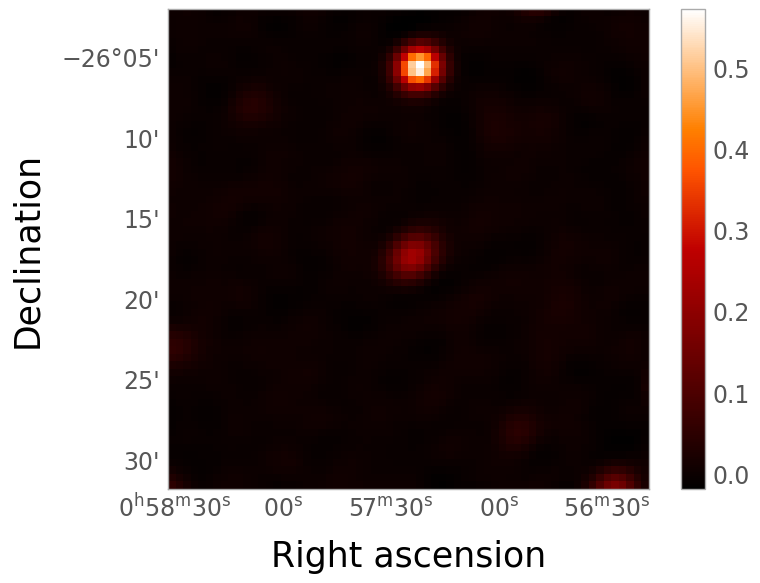}\\
\end{tabular}
\caption{An example of each type of 'halo' (in the central field of view, in the frequency range 170-231 MHz, with an angular extent of 29.86 arcmin $\times$ 29.86 arcmin) used in the dataset; From top left to bottom left in a clockwise direction - Abell 0141 (H), Abell 2693 (cH), Abell 0122 (cmH) and rxcj0137 (mH). The colour scales are all in units of Jy beam$^{-1}$}.
\label{fig:fig1} % I can do without the label too
\end{figure}

We use the GLEAM Survey, created wth the MWA to create our dataset. This survey includes observations of halos in different frequency ranges. The GLEAM survey covers frequency ranges of 72-103 MHz, 103-134 MHz, 139-170 MHz, and 170-231 MHz. This allows the ML model to learn the spectral shape of the halo source. As a result, the network can distinguish halos from Radio Galaxies and AGN, which do not have a steep spectral index \citep{feretti_2012,Gasperin_2017}. We create a dataset consisting of 23 radio image cubes, each with dimensions (300,300,4) in (RA, Dec, frequency), centered around sources that are classified as halo (H), candidate halo (cH), minihalo (mH), or candidate minihalo (cmH). For more details on their classification references and properties, refer to Table \ref{class_Halos}. Here $300 \times 300$ refers to the pixel size of the image along right ascension and declination. The angular extent of these images for the frequency ranges mentioned above are $4.667^{\circ} \times 4.667^{\circ}$, $3.667^{\circ} \times 3.667^{\circ}$, $2.833^{\circ} \times 2.833^{\circ}$, and $2.333^{\circ} \times 2.333^{\circ}$ respectively. In the dataset, we have 10 halos, 10 candidate halos, 1 minihalo, and 2 candidate minihalos. These all sources belong to the halo class for our binary classification problem. An example of these halo 'types' is shown in Fig. (\ref{fig:fig1}) in the form of central cut-outs of size $64 \times 64$ pixels with the same centre as the original images. 

We also prepare a dataset containing radio images of 23 non-halo sources which included normal radio galaxies, X-shaped radio galaxies, and some odd radio circles (ORCs). We refer to any radio galaxy without any unusual or exotic features as a normal radio galaxy. Thus, excluding the ORCs and X-shaped radio galaxies, the remaining non-halo sources consist of galaxy clusters with no reported halos (or central diffuse emission). We include ORCs and X-shaped galaxies in the dataset to develop a classifier capable of analyzing any radio observation and distinguishing diffuse emissions from these rare objects. Halos have smooth structures, unlike the complex shapes of X-shaped galaxies or ORCs, which could be misclassified as diffuse sources. Second, given the unknown nature of these objects, they require special attention and including them in any halo studies without proper classification introduces observational biases. Lastly, it diversifies the training data, helping the classifier distinguish between different sources and true diffuse emissions. We keep only 23 non-halo sources to create a class-balanced dataset in terms of sources. However, after augmentation, the balance between classes in the training set is disrupted, necessitating a modification of the loss function during training, as discussed in Section \ref{m2}.
The dataset is published on the MWA site and is collected on Github (\href{https://github.com/mishraashu6566/Radio-Halos-Classification-with-MWA.git}{github-radio-halo-classification}). This dataset is referred as \textit{GLEAM-Data} throughout the text and is used for our classification task. The radio images of the GLEAM survey are obtained from the virtual observatory skyview\footnote{\href{https://skyview.gsfc.nasa.gov/current/cgi/query.pl}{https://skyview.gsfc.nasa.gov/current/cgi/query.pl}}.
\begin{table}
    \centering
        \caption{Classification Table for all the 'halo' sources used in this paper}
    %\tiny
    \scriptsize
    \begin{tabular}{cccccc}
    \hline
    Cluster & z & Type & $\text{R.A.}_{J2000}$ (deg) & $\text{Dec}_{J2000}$ (deg) & Refs. \\
    \hline 
    Abell 0141 &  0.230 &  H  & 16.405 & -24.681  & (\hyperlink{m1}{a}) \\ 
    Abell 3404 &  0.164 &  H  & 101.373 & -54.227 & (\hyperlink{m1}{a}) \\
    Abell S1121 &  0.358 &  H & 351.299 & -41.204 & (\hyperlink{m2}{b})\\
    Abell S1063 & 0.347  & H and RG & 342.189 & -44.528 & (\hyperlink{m2}{b})\\
    MACS J2243.3-0935 & 0.447 & H & 340.838 & -9.595 & (\hyperlink{m2}{b}) \\
    Abell 2163 & 0.203 & H & 243.942 & -6.147 & (\hyperlink{m3}{c})\\
    Abell 2254 & 0.178 & H & 259.441 & 19.673 & (\hyperlink{m3}{c})\\
    Abell 2744 & 0.308 & H & 3.584 & -30.388 & (\hyperlink{m3}{c})\\
    PLCK G287.0+32.9 & 0.390 & H & 177.705 & -28.077 & (\hyperlink{m3}{c})\\
    RXC J1314.4-2515 & 0.247 & H & 198.606 & -25.262& (\hyperlink{m3}{c})\\
    Abell 2680 &  0.190 &  cH & 359.117 & -21.038 & (\hyperlink{m2}{b}),(\hyperlink{m4}{d})\\
    Abell 2693 &  0.173 &  cH & 0.537 & -19.555 & (\hyperlink{m2}{b}),(\hyperlink{m4}{d})\\
    Abell 2811 &  0.108 &  cH & 10.537 & -28.536 & (\hyperlink{m2}{b}),(\hyperlink{m4}{d})\\
    Abell 3186 &  0.127 &  cH & 58.086 &-74.003 & (\hyperlink{m4}{d})\\
    Abell 3399 &  0.203 &  cH & 99.326 & -48.484 & (\hyperlink{m4}{d})\\
    PSZ1 G287.95-32.98 &  0.250 &  cH & 74.849 & -75.802 & (\hyperlink{m4}{d})\\
    Abell 2496 &  0.123 &  cH (or cR) & 342.741 & -16.401 & (\hyperlink{m2}{b}),(\hyperlink{m4}{d})\\
    Abell 2721 &  0.117 &  cH (or cR) & 1.509 & -34.725 & (\hyperlink{m2}{b})\\
    Abell S0084 & 0.108 & cH & 12.347 & -29.526 & (\hyperlink{m2}{b})\\
    GMBCG J357.91841 & 0.394 & cH & 357.918 & -8.979 & (\hyperlink{m2}{b})\\
    Abell 0122 & 0.113 & cmH (or R) & 14.354 & -26.279 & (\hyperlink{m2}{b}),(\hyperlink{m4}{d})\\ 
    MCXC J0145.2-6033 & 0.180 & cmH & 26.299 & -60.563 & (\hyperlink{m2}{b}),(\hyperlink{m4}{d})\\
    RXC J0137.2−0912 & 0.039 & mH & 24.306 & -9.187 & (\hyperlink{m2}{b}),(\hyperlink{m4}{d})\\
    \hline
    \end{tabular}

\begin{minipage}{\textwidth}
        \scriptsize
        References (Refs.) : \hypertarget{m1}{(a)} \citet{Duchesne_2021}; \hypertarget{m2}{(b)} \citet{Stefan_2021}; \hypertarget{m3}{(c)} \citet{George_2017} \\(d) \citet{Duchesne_2021_low} ; z is the redshift\\
        Classification: Halo (H), Candidate Halo (cH), Relic (R), Mini-halo (mH), Candidate mini-halo (cmH), \\Individual or blended [remnant] radio galaxy (RG)\\
    \end{minipage}
    \label{class_Halos}
\end{table}   
The set of original images with sizes 300$\times$300 pixels are pre-processed using a procedure similar to \citet{Aniyan_2017}. First, we set all the pixel values below the $3\sigma$ level ($\sigma$ being the Root Mean Square (RMS) value of local noise in the image) of the background to remove contributions from any noise-dominated pixels. Ideally, a radio image is generated with a corresponding residual image, which we use to estimate the noise. The estimation is done by fitting a Gaussian to the histogram of flux values of the residual data. Due to the unavailability of the residual files for our GLEAM images, we used a slightly modified strategy to estimate the noise directly from the image. We fit a Gaussian to the lower end of the histogram of flux values, thus providing us with a noise estimate. To validate this method, we cross-checked some of the estimates manually using the following procedure: 
\begin{enumerate}
    \item Any five arbitrary patches in the image are chosen with almost no bright source emission or point sources that one can notice with eyes.
    \item Local noise RMS values are estimated for these patches using ds9 \citep{ds9}.
    \item The average of the values obtained is calculated to get the final estimate.
\end{enumerate}

\begin{figure*}
\centering
\hspace*{-1.0cm}
\begin{tabular}{cccc}
\includegraphics[width=3.8cm,height = 4.3cm]{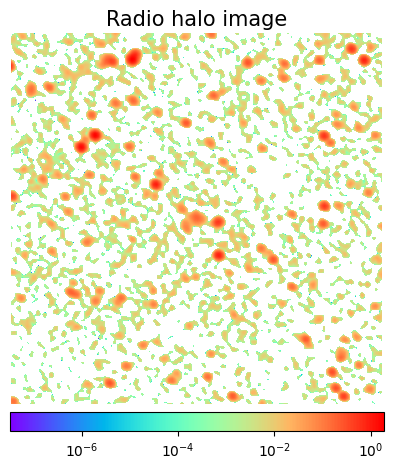} & \includegraphics[width=4.1cm,height = 4.3cm]{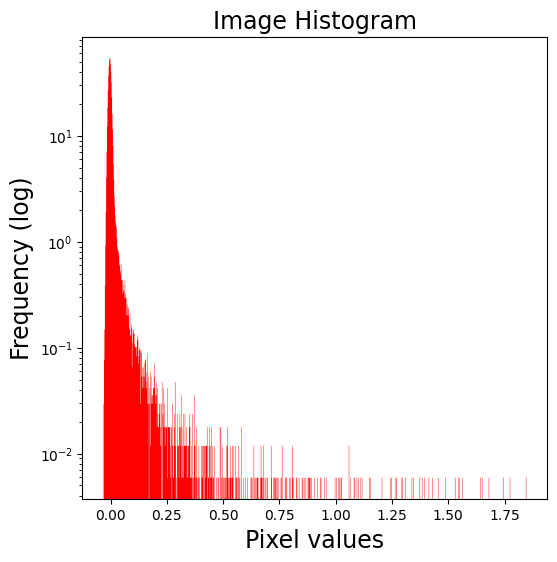}&
\includegraphics[height = 4.3cm,width=6.9cm]{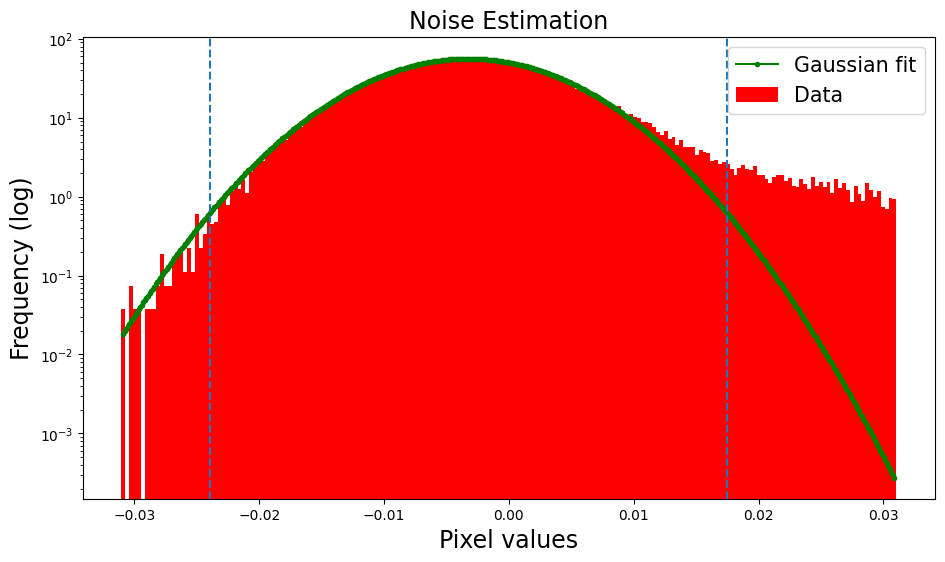}&
\includegraphics[width=3.5cm,height = 4.3cm]{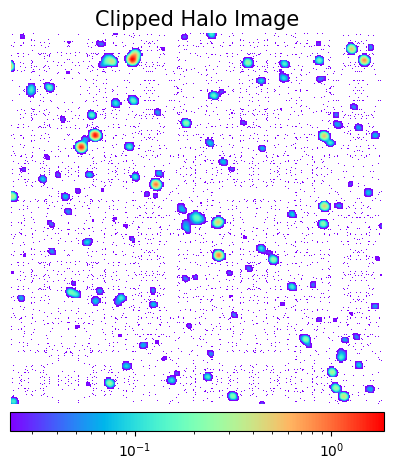}
\\
\end{tabular}
\caption{Pre-processing procedure for halo images. From left to right, the first panel shows an MWA-observed radio image. The second panel exhibits the image histogram with the salt-pepper background noise apparent towards the extreme left. The third panel of Figure 2 shows the full (red-filled histogram) distributions of the lower spectrum of the image histogram. Noise is estimated as $\sigma$ of this distribution. The dotted lines show the 3$\sigma$ boundaries. The fourth panel shows the final pre-processed Log-Normalized image where pixel values less than 3 times the local RMS noise are discarded. This removes most of the noise fluctuations in radio maps. The pixel values and the colour scales are all in units of Jy beam$^{-1}$.}
\label{fig:fig2} % I can do without the label too
\end{figure*}

\begin{figure}
    \centering
    \includegraphics[width =\linewidth]{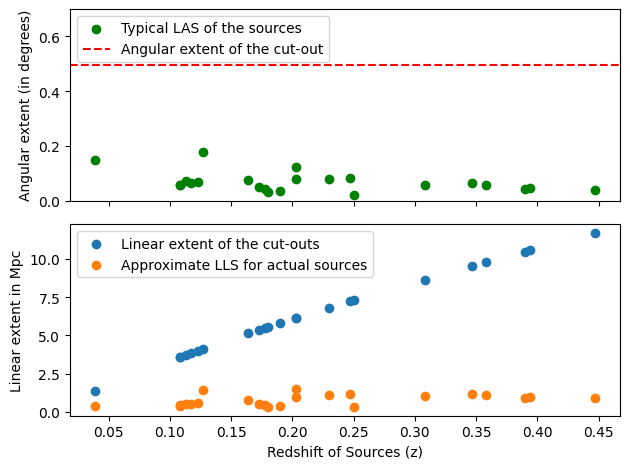}
    \caption{The top panel shows the halo sources' Largest Angular Size (LAS, green dots) against redshift, with the red dashed line marking the fixed angular extent of the 64x64 cut-out. The bottom panel compares the linear extent of the cut-outs (blue) to the sources' Largest Linear Size (LLS, orange). This confirms that the chosen cut-out size adequately contains all halo sources (For comparison, the cutouts with the smallest angular extents-frequency range 170-231 MHz-are selected). }
    \label{fig:extents}
\end{figure}

The entire procedure has been shown for one of the halos in Fig. (\ref{fig:fig2}). After RMS estimation of local noise, we generate 64 × 64 pixels central cut-outs for all images, using the same center as the original images. The cut-out size is chosen based on two key factors: the requirements for image generation and the extent of the halo sources. For image generation, smaller cut-outs generally result in better quality, but they cannot be too small, or they risk excluding the halos. Our analysis revealed that most halo sources in the central field of view could fit within a 32 × 32 pixels cut-out, but their extents near the boundaries would be truncated for a few. Therefore, a 64 × 64 pixels size proved ideal in our experiments, as it effectively captures the halos and allows efficient image generation. This is also verified from Fig. (\ref{fig:extents}) where the linear extents are calculated from the angular extents of the sources at given redshifts assuming $\Lambda$CDM cosmology (For typical sizes of the halo sources, we refer \citet{Cantwell_2016}, \citet{2022_MeerKAT}, and references from Table \ref{class_Halos}). Additionally, the extents of the cut-outs are evaluated within the frequency range of 170-231 MHz, as this range corresponds to the smallest angular extent among all considered frequencies.

These 64 $\times$ 64 cutouts are used for the image generation and classification task after performing the necessary normalization of the pixel values. There are few other sources present in these cut-outs and they can act as noise or distractors for the generative models. However, as radio observations with diffuse emission have naturally crowded field, the additional sources in the cut-outs can actually help the model to generalize better and learn to differentiate between different types of emission. It allows the model to recognize that the central halo is the primary target, while learning to handle the point sources as part of the background. In our case, having a crowded field means the generative model must account for and learn to deal with multiple sources entering the 64x64 pixels cut-outs, reflecting the real observational conditions. Similar arguments hold for our classification model as well. We rescale pixels to the range [0,1] for the classifier and to [-1,1] for image generation after log-scaling them. For more information about the normalization schemes, refer to the appendix \ref{NetA}.  

For generative models' comparison, we prepare an additional dataset with single-frequency images from the MWA survey, including 10 different sources (H, cH, and mH) observed with various Briggs weightings (\cite{Briggs_1995}; \cite{Boone_2013}). This dataset, also available on GitHub, undergoes the same pre-processing as the GLEAM-Data, allowing us to evaluate the generative models across diverse scenarios. It is worth noting that this additional dataset is just a benchmark to compare our generative models and does not contribute directly to our main results which employ the multifrequency GLEAM-Data set.

\section{Methods}\label{methods}
In this section, we describe our approach to enhancing halo classification performance through a combination of innovative techniques. We start by detailing the different classifier architectures we employ, including basic convolutional networks and those with advanced features such as attention mechanisms. We then discuss our data augmentation strategy, which includes both traditional techniques and novel generative methods using WGANs and DDPMs. This combination aims to address the challenges posed by limited dataset sizes and improve classification accuracy. Finally, we end the section by outlining details of the networks' training procedures.
 
\subsection{Halo Classifiers' Architecture}\label{Classif}
 
We develop various classifiers, each trained with different architectures on either augmented data by itself or augmented data supplemented with generated data to obtain the best classification performance. Our baseline is the classifier trained on augmented data alone. We also compare the classifier trained on original images plus generated images, with classifiers with different architectures trained on the original images. This is to assess the impact of architectural complexity compared to data augmentation in our classification task.

Our classification task is simple, involving two classes to predict. Class `1' are images with halos, while class `0' are non-halo images. The classifier outputs a "classification score" that ranges from 0 to 1, where it is trained to produce a score of 1 for halos and 0 for non-halos. This scoring mechanism allows the model to differentiate halos from other sources in the crowded field. The primary aim of our classifier is to differentiate between the halo and the `empty' space, as well as to distinguish the halo from other galaxies. We classify any side cut-out extracted from a halo image as an `empty' space. Therefore, our training and test sets include images from the non-halo class that depict either `empty' space or other galaxies. This is the standard setup for all our classifiers (Table \ref{tab:arch_lay}).

We begin with the straightforward LeNet-style architecture \citep{LeNet}, where the input is first processed by a series of convolutional layers for feature extraction, followed by fully connected layers for classification. In the convolutional layer, we use the ReLU activation function followed by the max-pooling layer. An overview of the architecture can be found in the appendix \ref{NetA}. This fundamental architecture is referred to as the base network throughout the paper and is denoted by I.

\begin{table}
    \centering
    \caption{Architecture Layout for Different Classifiers}
    \begin{tabular}{cccc}
    \hline
         & CNN (I)  & CNN with Attention (A)& Dense (D) \\
         \hline
       Layer 1  & Convolution 2D &  Convolution 2D & Dense \\
      Layer 2    & Max Pooling 2D &  Max Pooling 2D & \\
       Layer 3   & Convolution 2D &  Convolution 2D & Dense \\
       Layer 4   & Max Pooling 2D &  Max Pooling 2D & \\
       Layer 5 &  &  Convolution 2D & \\
      Layer 6  &  &  Multi-Head Functional & \\
        Layer 7  & Dense &  Dense & Dense  \\
       Layer 8  & Dense &  Dense & Dense \\
    \hline
    \end{tabular}
    \begin{tabular}{c}
 \{$\overline{\text{I}}$,$\overline{\text{A}}$,$\overline{\text{D}}$\} differs from \{I,A,D\} in terms of training data set.\\
    \hline
    \end{tabular}
    \label{tab:arch_lay}
\end{table}

We intend to keep the architecture of the classifier as general as possible to ensure it can detect diffuse extended emission, which may not typically be centrally located. Therefore, the network should be able to learn correlations between the pixels of the image at different positions. Further, we want the network to have a minimum number of trainable parameters to avoid overfitting. These reasons suggest the use of multiheaded attention to modify the architecture of the conventional convolutional network. This network is denoted by A. Finally, a simple classifier network with fully connected layers is denoted by D. The architecture is tabulated in the appendix \ref{NetA}.

\subsection{Data Augmentation and Training}\label{m2}

An 80-20 train-test split is performed on the original dataset of the halo sources (H, cH, mH, cmH) based on the number of available sources before applying any augmentations. This ensures that the training and validation datasets remain blind to the augmentations, preserving the integrity of the test set. We have only 23 sources in the halo class (i.e., class '1'). Compared to standard data sets of radio sources, such as MiraBest\footnote{The MiraBest dataset can be downloaded from: \href{https://doi.org/10.5281/zenodo.4288837}{https://doi.org/10.5281/zenodo.4288837}.} \citep[publicly available Fanaroff-Riley (FR)-labeled ML dataset of radio galaxies;][]{2017_MiraBest}, there are fewer sources in our study. 
So, we apply extensive image augmentation to create a reasonable training and test dataset. In addition to standard image augmentation, that is, rotations and flips, we also employ pixel scaling by taking scaling coefficients randomly from the range [0.9,1.4]. We select multiples of 45 degrees from 0 to 360 degrees range to rotate the images and perform horizontal and vertical flips. This means 8 ways of rotation, 2 ways of flips, and 10 ways of performing brightness scaling. As a result, each image gives $8\times 2 \times 10 $ augmented samples. The prepared dataset details are listed in Table \ref{tab:train-test}. Initially, we also tried size scaling but the network does not perform well with such augmentation in generating images. More experiments are required to assess the performance of the classifier with size scaling as an additional augmentation.
\begin{table}
\centering
\caption{Data used in this work: The table shows the number of sources in
each class that are provided in the training and test partitions for the GLEAM
data set, along with the number of images used after augmentation (both classic and generative) }
\begin{tabular}{lcccc}
\toprule
\textbf{Data} & \multicolumn{2}{c}{\textbf{Train}} & \multicolumn{2}{c}{\textbf{Test}}  \\
\cmidrule(lr){2-3} \cmidrule(lr){4-5}
 & Class 1 & Class 0 & Class 1 & Class 0   \\
 \hline
\textit{GLEAM Sources} & 18 & 18 & 5 & 5 \\
\textit{Classic Augmentation} & 5440 & 5200 & 1280 & 960\\
\textit{Classic+Generative} & 11264 & 5200 & 1280 & 960\\
\bottomrule
\end{tabular}
\begin{tabular}{c}
Input image size is (64, 64, 4), where 4 denotes frequency channels.\\
\hline
\end{tabular}
\label{tab:train-test}
\end{table}
To further increase our classifiers' performance, we explore generative data augmentation. We employ WGANs and DDPMs described in Section \ref{bkgd} to generate halo images. Initially, we train each model on single-frequency images of 10 different halos as described in Section \ref{Dat} to determine which performs better. We evaluate the models using a range of metrics to assess their effectiveness. It turns out DDPMs perform better across all metrics. So, DDPM is utilized to generate multi-frequency images of halos from our training dataset of the classifier. To enhance the training of the generative models and prevent overfitting, Gaussian noise with 0 mean and very small standard deviation, was used to augment the training set, thus increasing its size and improving model robustness. These generated images are subsequently used to enhance the performance of the radio-halo classifier. The classifier networks \{I,A,D\} when supported with diffusion-based image augmentation are referred as \{$\overline{\text{I}}$,$\overline{\text{A}}$,$\overline{\text{D}}$\}. 

We train our generative models using 2 NVIDIA V100 PCIe 32GB GPUs (2×7 TFLOPS) on the IZAR cluster\footnote{\href{https://www.epfl.ch/research/facilities/scitas/hardware/izar/}{https://www.epfl.ch/research/facilities/scitas/hardware/izar/}} (Hypercomputing server available at SCITAS).  All these models are implemented with Tensorflow \citep{tensorflow_2016}. Optimization of these models is performed with an Adam optimizer \citep{kingma2017adam}, and a scheduler is used to dampen the learning rate in DDPMs. For more details regarding the hyperparameters of the training, refer to appendix \ref{NetA}.

The classifier networks are trained for 15 to 25 epochs using a class-balanced categorical cross-entropy loss with the Adam optimizer. This class-balanced loss function addresses the class imbalance in the training data following augmentation. All subsequent results are generated using this class-balanced loss. The default learning rate (Appendix \ref{NetA}) is used with a batch size of 128. No additional hyperparameter training is performed. Early stopping is used to prevent overfitting, and the checkpoint stores the model based on the validation accuracy. All classifiers are trained and evaluated in TensorFlow.

\section{Results and Discussions} \label{ress}
Here, we present our results. We begin by evaluating generative models trained on single-frequency images and identifying DDPMs as the best model based on generated distributions, images, and FID and KID scores. We then use DDPM to generate multi-frequency images and assess their quality using the same set of metrics. We then present our classifiers' convergence and performance metrics, showing that diffusion-assisted classifiers outperform others, with the attention-based classifier achieving the highest validation accuracy. After that, we apply the improved classifier to search for halos in GLEAM observations using various cluster catalogs, listing the sources our classifier successfully detects. The section concludes with a discussion of our findings.
\subsection{Evaluation of Generative Models trained on Single-Frequency images}\label{Res1}
We assess the quality of our generated samples through a comprehensive three-fold evaluation. Initially, we analyze histograms of pixel values from real and generated images, comparing the relative error per bin. A visual inspection of the images follows this, and finally, we compute standard evaluation metrics for the generated samples. This approach is consistently applied to both WGANs and DDPMs trained on single-frequency images.

Our results indicate that DDPMs significantly outperform WGANs, demonstrating superior performance in approximating the true flux distribution with reduced relative error per bin and producing higher-quality images with less noise (see Fig. (\ref{Gen_app1}) and Fig. (\ref{Gen_app2}) in the appendix \ref{GM_Res}). Quantitatively, DDPMs achieve lower FID and KID scores of 8.40 and 2.25, respectively, indicating higher-quality generated samples than WGANs. A detailed summary of these results is provided in Table \ref{tab:SM_res}, which supports the use of DDPMs for multi-frequency halo image generation.
\begin{table}
    \centering
    \begin{tabular}{|c|c|c|c|}
    \hline
        Models &  FID (10k samples) & KID  \\
    \hline
         WGAN  & 25.02 & 2.85  \\ 
    \hline
        DDPM & \textbf{8.40} & \textbf{2.25} \\ 
    \hline
    \end{tabular}
    \caption{Evaluation metrics (Fréchet Inception Distance and Kernel Inception Distance) for generated Halo images for cWGAN and DDPM}
    \label{tab:SM_res}
\end{table}
\subsection{Multi-Frequency Halo Image Generation with DDPM}\label{sec_MF}
Having determined that DDPM outperforms WGAN on relatively simple single-frequency halo images, we use DDPMs to generate multi-frequency halo images from GLEAM observations. The main difference from single-frequency image generation lies in the data used for training and a minor adjustment in the DDPM architecture—specifically, modifying the channel numbers in the final convolution layer to match the different frequency ranges of the GLEAM images. The training dataset consists of 5,440 images from 18 halo sources (the same set used for training classifiers), with each image having dimensions of (64, 64, 4), where 4 corresponds to the frequency channels.  The generated images are tested with respect to the same training set, as is the standard practice.

\begin{figure*}
\centering
\hspace*{-0.5cm}
\begin{tabular}{cc}
\includegraphics[height = 7cm,width=0.5\linewidth]{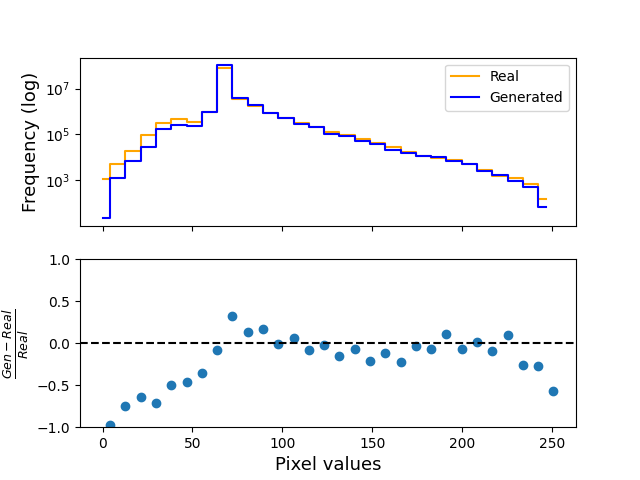}&
\includegraphics[height = 7cm,width=0.5\linewidth]{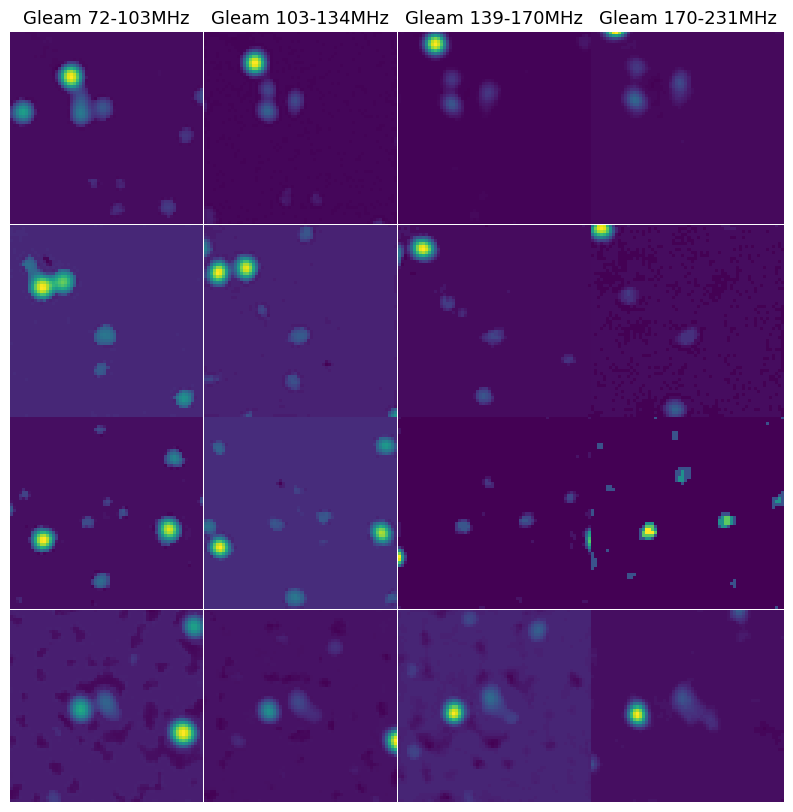}
\end{tabular}
\caption{Diffusion Model Evaluation: Pixel value distributions for diffusion-generated (blue) and real multi-frequency (orange) halo images are shown, with per-bin relative error in the bottom left panel. On the right, augmented unconditional diffusion-based images are displayed for multi-frequency image generation.}
\label{MF_IMG_GEN}
\end{figure*}

Figure (\ref{MF_IMG_GEN}) shows a pixelwise image distribution analysis and samples of arbitrarily chosen images, indicating a good approximation of the true flux distribution (overall) with minor challenges in accurately recreating low but non-zero flux values. The KID for these images, calculated with a batch size of 32, is 1.04, and the FID for 10,000 samples is 24.51. These generated images are used to augment the training dataset for the halo classifier.

\subsection{Convergence and Performance of Classifiers}

\subsubsection{Convergence}
\begin{figure*}
\centering
\hspace*{-0.5cm}
\begin{tabular}{cc}
\includegraphics[width=0.5\linewidth]{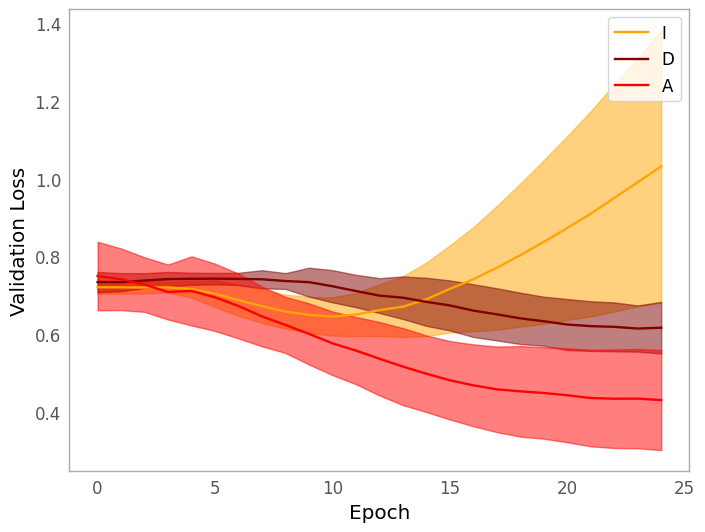}&
\includegraphics[width=0.5\linewidth]{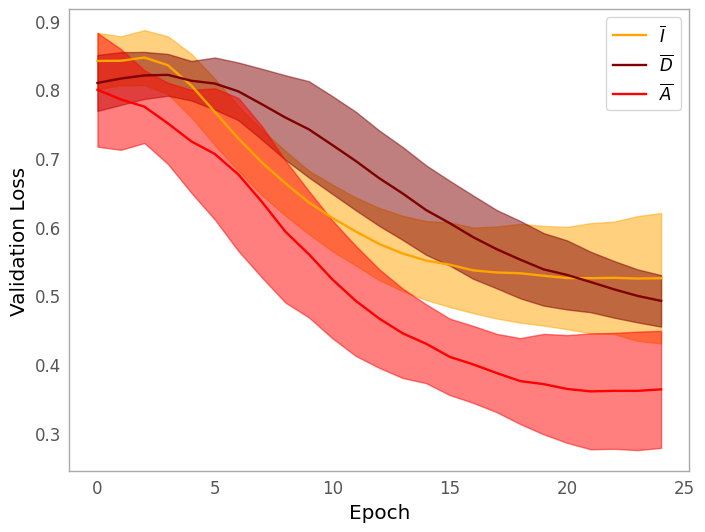}
\end{tabular}
\caption{Validation losses during the training of (i) the base classifier \{I, A, D\} (Table \ref{tab:arch_lay}; Appendix \ref{NetA}) models for the GLEAM-Data set (left), and (ii) their diffusion-assisted counterparts  \{$\overline{\text{I}}$,$\overline{\text{A}}$,$\overline{\text{D}}$\} for the GLEAM-Data set (right). Plots show mean and standard deviation over 10 training runs where the losses are smoothed with an exponential moving average with $\alpha$ = 0.1 to remove small-scale variability}
\label{Class5}
\end{figure*}

Validation loss curves for both the base networks \{I, A, D\} and their diffusion-assisted counterparts are shown in Fig. (\ref{Class5}). Curves depict the mean and standard deviation for each network over 10 training runs. It can be seen from Fig. (\ref{Class5}) that the standard CNN I suffers from overfitting very early in training. The attention-based classifier supported by diffusion-based augmentation, $\overline{\text{A}}$ performs best in minimizing the validation loss. We see an improved validation loss for each diffusion-assisted classifier \{$\overline{\text{I}}$,$\overline{\text{A}}$,$\overline{\text{D}}$\} over its corresponding base network \{I, A, D\}.

The improved performance of the attention-based networks A and $\overline{\text{A}}$  can be attributed to the significant decrease in the learnable parameters compared to I and $\overline{\text{I}}$, and their ability to learn pixel correlations across any region within the image. In contrast, the convolutional layers in I and $\overline{\text{I}}$ can only retain the local, and global information within their fixed receptive fields.
The next noticeable thing from Fig. (\ref{Class5}) is that the stability of validation loss for \{$\overline{\text{I}}$,$\overline{\text{A}}$,$\overline{\text{D}}$\}, in comparison to base classifiers \{I, A, D\}, indicates that image data augmentation contributes to stabilizing the training process and preventing overfitting. Classifier I clearly shows signs of overfitting in Fig. (\ref{Class5}) for a larger number of epochs. Therefore, the use of regularization techniques is essential for classifier I if it is to be trained for more than 10 epochs without early stopping in future experiments. 

\subsubsection{Performance: metrics}
Standard classification metrics for both the base networks \{I, A, D\} and their diffusion-assisted counterparts used are shown in Table \ref{tab:res}. The metrics in this table are calculated using the held-out test set from our GLEAM Halos dataset, classified using the best-performing model as per the stored checkpoint based on the maximum validation accuracy criterion. The values in the table show the mean and standard deviation for each metric over 10 training runs.
From Table \ref{tab:res}, it can be seen that the best test accuracy is achieved by model $\overline{\text{A}}$, highlighted in bold. Further, we see that the diffusion-assisted model performs better than its base network. 
\begin{table*}
    \centering
    \caption{Performance metrics for classification of the Gleam Halo data set using the classifiers \{I, A, D\} (Table \ref{tab:arch_lay}; Appendix \ref{NetA}) and their corresponding diffusion-assisted counterparts \{$\overline{I}$, $\overline{A}$, $\overline{D}$\}}
    \begin{tabular}{ccccc}
    \hline
    \hline
    \textbf{GLEAM-Data} & \textbf{Accuracy [\%]} & Precision & Recall & F1-Score \\
    \hline
    I & 86.87 $\pm$ 2.78 & 0.905 $\pm$ 0.014 & 0.860 $\pm$ 0.052 & 0.881 $\pm$ 0.029 \\
    D & 84.68 $\pm$ 1.03 & 0.869 $\pm$ 0.019 & 0.863 $\pm$ 0.033 & 0.865 $\pm$ 0.011\\
    A & 94.41 $\pm$ 1.95 & 0.934 $\pm$ 0.032 & 0.972 $\pm$ 0.017 & 0.952 $\pm$ 0.016 \\
    \hline
    $\overline{\text{I}}$ & 92.05 $\pm$ 2.94 & 0.942 $\pm$ 0.010 & 0.918 $\pm$ 0.058& 0.929 $\pm$ 0.029 \\
    $\overline{\text{D}}$ & 90.45 $\pm$ 0.84 & 0.914 $\pm$ 0.012 & 0.920 $\pm$ 0.015& 0.917 $\pm$ 0.007 \\
    $\overline{\text{A}}$ & \textbf{95.93} $\pm$ \textbf{1.57} & 0.962 $\pm$ 0.017 & 0.967 $\pm$ 0.016& 0.965 $\pm$ 0.014 \\
    \hline
    \end{tabular}
    \label{tab:res}
\end{table*}
For a binary classification problem with true positives $TP$, false positives $FP$, and false negatives $FN$, the precision is defined as $TP/(TP+FP)$ and the recall is defined as $TP/(TP+FN)$.
Thus, recall represents the proportion of the true positives correctly identified by the classifier. The harmonic mean of these two metrics, precision and recall, is called the F1-Score \citep{vanrijsbergen1979information,james2013introduction}.
\begin{figure}
    \centering
    \includegraphics[width = \linewidth]{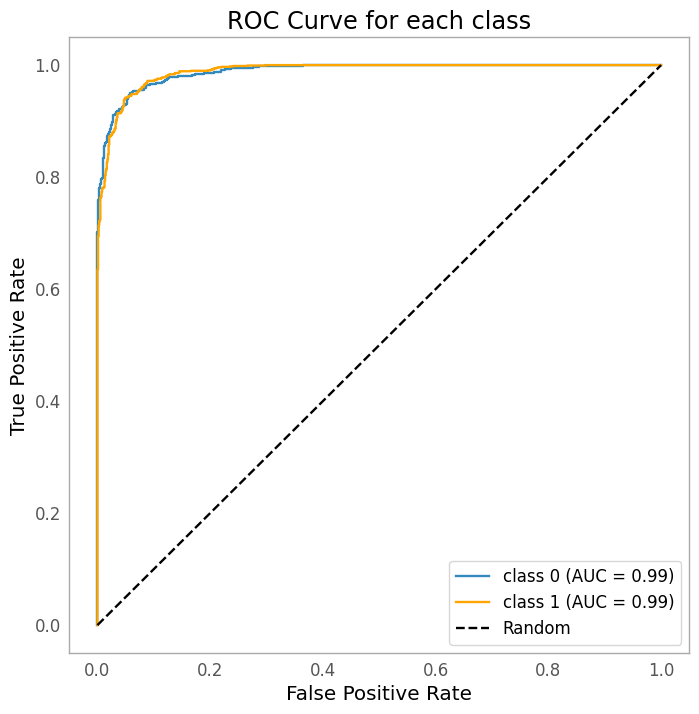}
    \caption{Receiver Operating Characteristic (ROC) Curves (true positive rate against the false positive rate) for Classifier $\overline{\text{A}}$ (Table \ref{tab:arch_lay}; Appendix \ref{NetA}) for one of the arbitrarily chosen training runs; Class 0 and Class 1, represent 'halos' and 'non-halos' respectively} 
    \label{Class3}
\end{figure}
As shown by the F1-scores in Table \ref{tab:res}, the attention-based model (F1-score of $0.952 \pm 0.016$) outperforms the conventional CNN (F1-score of $0.881 \pm 0.029$).  In Fig. (\ref{Class5}), it is evident that the attention model (A) exhibits higher validation accuracy than the CNN (I). Thus, using the same CNN structure and incorporating attention results in a reduction of trainable parameters and improved validation accuracy, as seen in our experiments.  Similarly, as expected the Diffusion-aided classifier performs better than the base network. In particular, $\overline{\text{A}}$ exhibits excellent performance with an average validation accuracy of 
95.93\% and F1 Score of $0.965 \pm 0.014$.

\begin{figure}
    \centering
    \includegraphics[width = \linewidth]{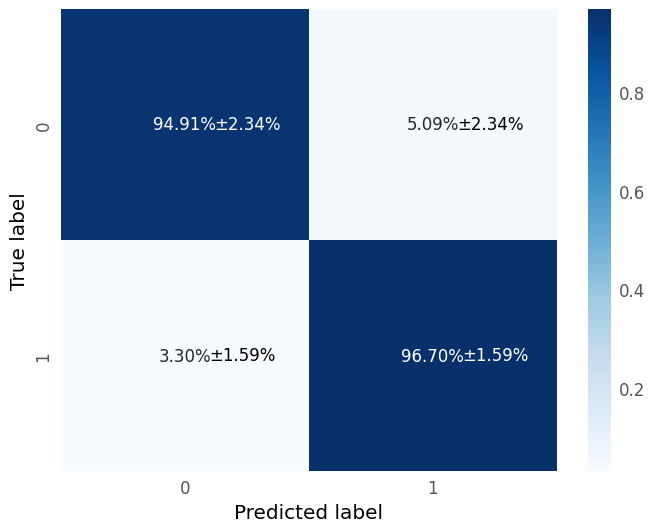}
    \caption{Confusion Matrix normalized over rows for Classifier $\overline{A}$ (Table \ref{tab:arch_lay}; Appendix \ref{NetA})): The matrix shows the mean and standard deviation of the original entries of the matrices over 10 training runs; Class 0 and Class 1, represent 'halos' and 'non-halos' respectively}
    \label{Class4} 
\end{figure}

Further, we observe in detail one of the best-performing models for $\overline{\text{A}}$ using additional metrics. The Receiver Operating Characteristic (ROC) curves for the halo and non-halo classes are depicted in Fig. (\ref{Class3}), showing the true positive rate (TPR) plotted against the false positive rate (FPR) for each class. An ideal classifier has a TPR of 1 and an FPR of 0; practically, a high TPR with a very low FPR indicates a 'good' classifier. The normalized confusion matrix over the rows (averaged over 10 training runs) is shown in Fig. (\ref{Class4}), where ideally, the diagonal boxes should be 100\% and the non-diagonal elements 0\% with 0\% uncertainty. Given this, the attention model's results are very good.

\subsection{Search For Halos in GLEAM Dataset using Various Cluster Catalogues}
Using our classifier network, we investigate sources listed in various cluster catalogues within the GLEAM datasets. Evaluating these sources, which were not included in the training or test sets, is essential for assessing the generalizability and robustness of our halo classifier. This approach validates the classifier's performance across diverse datasets, identifies new potential halos, and helps refine the model by addressing any limitations. 

In this study, we first search for potential halos in MGCLS and PSZ2-discovered diffuse sources to test our classifier’s performance. 
We only consider a source to be a halo if the classification score $\geq$ 0.95 to minimize false positives and exclude sources where instrumental artifacts or noise might lead to misclassification. We then apply the classifier to completely unlabeled sources from the XMM-Chandra catalogue to identify new potential halos. Analysis of these results to refine the classifier’s limitations is deferred to future work.

\subsubsection{MGCLS and PSZ2-based Halos Rediscovered in GLEAM}
We use the MGCLS diffuse sources classification performed in \citet{2022_MeerKAT}. We obtain their corresponding GLEAM observations. After performing necessary preprocessing as outlined in Section \ref{Dat}, the images are passed through our trained classifier, which was not exposed to these sources during training. Notably, these sources were not identified in the GLEAM catalogue \citep[][]{Stefan_2021, Duchesne_2021_low, George_2017}, yet our classifier successfully detects them (9 out of 12) with high accuracy, as indicated by a classification score exceeding 0.95. This demonstrates the classifier’s potential for automated detection of previously unrecognized halos, as listed in Table \ref{tab:MGCLS}.
\begin{table}
    \centering
    \begin{tabular}{cccc}
    \hline
    \hline
    Cluster Name &  $\text{R.A.}_{J2000}$ (deg)  & $\text{Dec}_{J2000}$ (deg) & Alternate name\\
    \hline
    Abell 0370 & 39.960 & −1.586 & ZwCl 0237.2−0146\\ 
    Abell 2667 &  357.920 & −26.084 & MCXC J2351.6−2605 \\
    Abell 0209 & 22.990 & -13.576 & MCXC J0131.8−1336\\ 
    Abell 0521 &  73.536 & -10.244 & MCXC J0454.1−1014 \\
    Abell 0545 &  83.102 & -11.543 & MCXC J0532.3−1131 \\
    Abell 3558 &  201.978 & -31.492 & MCXC J1327.9−3130 \\
    Abell 3562 &  202.783 & -31.673 & MCXC J1333.6−3139 \\
    J0516.6-5430 &  79.158 & -54.514 & Abell S520 \\
    J0638.7-5358 &  99.694 & −53.972 & Abell S592 \\
        
    \hline
    \end{tabular}
    \caption{List of MGCLS halos re-discovered in Gleam Observations using Diffusion-assisted Attention Classifier $\overline{\text{A}}$ : Only sources with classification scores greater than 0.95 have been considered}
    \label{tab:MGCLS}
\end{table}

We also investigate halos from the PSZ2 catalogue that were detected in the LOFAR sky survey \citep[][]{2022_Botteon}. Out of 83 halo sources identified in the PSZ2 catalogue, only 17 are found in the GLEAM dataset for whom preprocessing is possible due to the limited sky overlap between GLEAM and LOFAR survey. GLEAM primarily covers the Southern Hemisphere, while LOFAR focuses on the Northern Hemisphere, restricting the number of halo sources from the PSZ2 catalogue detectable in GLEAM. These 17 sources are analyzed using the same procedure as for MGCLS sources with our classifier. We successfully rediscovered five of these halos, which are listed in Table \ref{tab: PSZ2}.
Of the remaining 12, we found that 3 consisted of truncated images (an image whose certain portion is entirely filled with 0, indicating the absence of data from the instrument) unsuitable for classification, and 4 had high noise levels compared to our training data (rms values on the order of 1 mJy beam$^{-1}$ or higher) across the frequency ranges. Additionally, 2 were smaller-sized halos relative to the typical halo sizes in our training set. For the rest of the three, we did not find any straightforward reason for their misclassification. A multi-frequency analysis of those sources will be required to indicate a concrete reason. Still, this detection rate (5/8 halos) is promising, given that the images used in this analysis are from GLEAM, which has a much coarser resolution compared to the LOFAR sky survey, where resolutions can reach a few arcseconds. Furthermore, LOFAR offers superior sensitivity and a lower typical rms noise of 0.1 mJy beam$^{-1}$\citep[][]{2022_Botteon}. 
\begin{table}
\centering
\begin{tabular}{cccc}
    \hline
    \hline
    Cluster Name & $\text{R.A.}_{J2000}$ (deg) & $\text{Dec}_{J2000}$ (deg) \\
    \hline
    PSZ2 G141.05-32.61 & 30.071 & 27.806  \\
    PSZ2 G031.93+78.71 & 205.461 & 26.353  \\
    PSZ2 G077.90-26.63 & 330.220 & 20.974  \\
    PSZ2 G116.50-44.47 & 8.034 & 18.107  \\
    PSZ2 G123.00-35.52 & 12.911 & 27.333 \\
    \hline
\end{tabular}
\caption{List of PSZ2 Detected halos re-discovered in GLEAM Observations using Diffusion-assisted Attention Classifier $\overline{\text{A}}$ with classification scores greater than 0.95}
\label{tab: PSZ2}
\end{table}
\subsubsection{COSMOS Field in GLEAM using XMM-Chandra Catalogue}
Like MGCLS and PSZ2, we search the COSMOS field as well for halos using the XMM-Chandra cluster Catalogue. These images are also preprocessed before they are passed on to the trained classifier for classification scores. The threshold for halo consideration is 0.95 and the sources matching the criterion are listed in Table \ref{tab:COSMOS}.
This demonstrates the classifier's capability to identify halo candidates in diverse datasets, highlighting its potential for broad application. However, the effectiveness of the classifier may be influenced by preprocessing quality, dataset variations, and limitations inherent in the training data. These factors underscore the need for further analysis to validate and refine the classifier's performance and the sources' labeling.
\begin{table}
    \centering
    \begin{tabular}{ccc}
    \hline
    \hline
    Cluster Name &  $\text{R.A.}_{J2000}$ (deg)  & $\text{Dec}_{J2000}$ (deg) \\
    \hline
    CXOC J100042.3+014534 & 150.177 & 1.760 \\ 
    CXOC J100030.4+023735 &  150.127 & 2.627  \\
    CXOC J100024.9+023956 & 150.104 & 2.666  \\
    CXOC J100022.7+023801 &  150.095 & 2.634  \\
    CXOC J100024.6+023748 &  150.103 & 2.630  \\
    CXOC J100043.1+014608 &  150.180 & 1.769 \\
    CXOC J100031.1+023729 &  150.130 & 2.625  \\
    CXOC J100045.5+014627 &  150.190 & 1.774  \\
    CXOC J095800.9+015757 &  149.504 & 1.966 \\
    CXOC J100045.9+014819 & 150.191 & 1.805 \\
    CXOC J100031.2+023642 &  150.130 & 2.612  \\

    \hline
    \end{tabular}
    \caption{List of XMM-Chandra Clusters classified as 'halos' in Gleam Observations using Diffusion-assisted Attention Classifier $\overline{\text{A}}$: Only sources with classification scores greater than 0.95 have been considered}
    \label{tab:COSMOS}
\end{table}

\subsection{Discussions}\label{disc} 
\subsubsection{Generative Models Assisted Classifiers}
In our study, halo images are generated to supplement the data set for our radio-halo classifier, given the limited number of labeled halos in MWA. \citet{Kummer_2022} and \citet{rustige_2023} used WGAN-generated images of FR I, FR II, Compact, and Bent radio galaxies to enhance classifiers for four-class classification tasks. They found that Dense and Convolutional architecture classifiers benefited the most, while complex architectures like the Vision Transformer \cite[ViT;][]{2020_ViT} saw less improvement. They also observed that the results depended on the number of generated samples used for augmentation.

In our case, the goal was not to achieve robust comparison among classifiers but to use diffusion-supported classifiers to optimize halo classification. Our classifiers are simpler in architecture, and thus we observed improvements across all of them due to the additional augmentation. Unlike \citet{rustige_2023}, we kept the number of generated samples constant and used diffusion-generated instead of GAN-generated samples, making direct comparison challenging. However, similar to their findings, we observed a decrease in improvement as the architectural complexity of classifiers increased, as shown in Table \ref{tab:res} and Fig. (\ref{Class5}). One possible reason is that complex networks can learn finer details from the training dataset that simpler networks might miss due to pooling, fewer layers, and other factors. When given generated samples, these complex architectures may end up learning irrelevant details. However, we find in agreement with the previous study that the statistical prowess of the training set increases across architectures, facilitating stable training and delayed onset of overfitting even in absence of any regularization techniques.

There is limited research employing diffusion models for image generation or reconstruction in radio astronomy. \citet{2023_RADiff} shares our motivation, using Latent Diffusion Models \citep[LDMs;][]{LDM} with conditional generation trained on annotated radio datasets to create synthetic images for augmenting datasets and reducing class imbalance in deep networks. Our work differs in that our dataset is smaller, allowing us to use DDPMs instead of LDMs without concern for inference timing. However, LDMs present an interesting prospect for future work with MGCLS.

\subsubsection{Note on regularization techniques and hyper-parameter tuning}
We observe an increase in the validation accuracy of the classifiers with the diffusion-aided augmentation scheme (see Table \ref{tab:res}). However, the ratio $\Gamma =  \frac{N_g}{N_r} \approx 0.5$ ($N_g$ and $N_r$ being the number of generated and real samples in the training set, respectively) remains constant for us. Our goal was to achieve the highest accuracy for the radio-halo classifier. Therefore, we did not extensively adjust this parameter, particularly given the slow generation of diffusion samples. Thus, it remains a hyperparameter for future tuning. \citet{rustige_2023} found that the performance worsens with an increased number of generated samples in the training dataset, as the network learns artifacts from the generative model rather than the actual properties of real images. It would be interesting to see how this parameter affects our classifiers' performance in follow-up work using LDMs and MAMBA-based diffusion models \citep[][]{Mamba_2024}, as hyperparameter tuning might lead to further improvements.

Additionally, we did not apply any regularization techniques in our networks, except for the attention-based classifier, which included a Gaussian noise layer. This was intentional to gauge the regularizing effects of diffusion-aided augmentations. We find that these samples do indeed regularize the network, as evident in Fig. 
 (\ref{Class5}).

\subsubsection{Sources detected as "halos" in cluster catalogs}
Table \ref{tab:MGCLS} lists diffuse sources classified as halo, minihalo, or candidates that we rediscovered in the GLEAM dataset. This was unexpected given the number of sources recovered, likely due to the different resolution of the GLEAM survey compared to MGCLS. We believe this effectiveness is rooted in our classifiers' ability to learn the properties of halos across different frequencies and thereby varying resolutions. Further, Table \ref{tab:COSMOS} lists the sources our trained classifier identified as 'halos' in the COSMOS field. We did not find counterparts for most of these in the MIGHTEE survey, but confirming them as halos will require further evidence and analysis. A key limitation of our dataset is that over half of the sources are classified as candidates, making it more suited for detecting central diffuse emissions rather than confirmed radio halos. Additionally, the MWA's poor angular resolution and lower signal-to-noise ratio restrict the variety of halos available for training, potentially reducing the classifier's ability to generalize to smaller or fainter halo structures.
So, this classification should be seen as the identification of a central source of diffuse emission in the cluster. Further analysis is necessary to provide more detailed labeling.
We also list the sources detected in the PSZ2 catalogue which were not part of the LOFAR sky study in the GitHub\footnote{\href{https://github.com/mishraashu6566/Radio-Halos-Classification-with-MWA.git}{github-radio-halo-classification}} which will help astronomers to narrow down their search. We encourage radio astronomers to analyze these further.

\section{Conclusion}\label{CON}
In this paper, we developed a classifier, unbiased to cluster selection methods, to identify diffuse emissions in MWA observations and successfully used it to rediscover several halos listed in the MGCLS and PSZ2 Catalogue within the GLEAM datasets. Additionally, we have demonstrated that the use of generative models, specifically diffusion, to augment the dataset results in a performance improvement over the base network (subject to architecture) for the radio halo classification problem considered here, without further hyperparameter tuning and regularizations. We have demonstrated that both the attention network in comparison to the traditional CNN, and the diffusion-assisted classifier compared to the base network, converge to higher validation accuracy values. These improvements can be attributed to two main factors. First, the attention's ability to capture correlations in the image at different positions. Second, the general ability of NNs to perform well when faced with more data.

Among the set of classically augmented classifiers, the attention-based classifier (A) delivers the best performance. Although diffusion-based augmentations enhance the performance of a simple network used as a classifier, they are of marginal importance for more sophisticated architectures, such as attention-based models. It should be noted that the results presented here are still preliminary and indicative of the trends that, in general, one can expect with the employment of these methods. For a robust comparison among the classifiers, k-fold validation needs to be implemented which we leave for our future work. In principle, WGAN samples can also be useful in augmenting the dataset despite its sub-par quality, and this has to do with the nature of the task at hand, that is, classification. However, the use of such samples for more sophisticated and sensitive tasks, such as parametric inference, is not recommended, as the contours are highly sensitive to the emulator samples.

Further, the high validation accuracy of the classifiers obtained here is attributed to two reasons: 1) the classification is binary (halos and non-halos) and 2) the dataset, both training and validation, is quite simple in the sense that the non-halos included are either cutouts or normal parts of galaxy clusters. The validation accuracy is expected to decrease if the validation set becomes more diverse and complex by including exotic species and incorporating multi-survey images of the sources. Nevertheless, we believe that the validation accuracy can still exceed 90\% with the aid of image augmentations from generative models like WGAN and stable diffusion, as well as by utilizing better networks such as multi-modal neural networks.

However, in conclusion, we echo the expectation that image augmentation with generative models will soon become the common choice, especially in astronomical problems where the labeled datasets are rare and limited. Radio halo classification is one such problem and even then our models and tools are still to be applied to high-resolution datasets such as MGCLS. Further work needs to be done in this direction, especially concerning the extension of the halo classification directly from the u-v plane rather than the image plane for more generality.
\section*{ACKNOWLEDGEMENTS}\label{qlock acknowledgement}
AK acknowledges the financial support from the EPFL Laboratory of Astrophysics (LASTRO) during the internship.  SK acknowledges the financial support from the SNSF under the Sinergia Astrosignals grant (CRSII5\_193826). ET acknowledges travel \& collaboration support from the SNSF under the Weave/Lead Agency RadioClusters grant (214815). The authors thank Stefan Duchesne for his valuable discussions during the project. This work was supported by EPFL through the use of the facilities of its Scientific IT and Application Support Center (SCITAS). The use of facilities of the Swiss National Supercomputing Centre (CSCS) is gratefully acknowledged.
\section*{DATA AVALAIBILITY}
Code for this work is publicly available on Github at the following address: (\href{https://github.com/mishraashu6566/Radio-Halos-Classification-with-MWA.git}{github-radio-halo-classification}). 
The GLEAM dataset is available on \href{https://skyview.gsfc.nasa.gov/current/cgi/query.pl}{the virtual sky observatory} and can be extracted using Astroquery python package\footnote{\href{https://astroquery.readthedocs.io/en/latest/}{https://astroquery.readthedocs.io/en/latest/}}.
\bibliographystyle{mnras}
\bibliography{main} % if your bibtex file is called example.bib

%\begin{thebibliography}{99}
%\end{thebibliography}

%%%%%%%%%%%%%%%%%%%%%%%%%%%%%%%%%%%%%%%%%%%%%%%%%%

%%%%%%%%%%%%%%%%% APPENDICES %%%%%%%%%%%%%%%%%%%%%

\begin{appendices}
\section{NORMALIZATION, AND NETWORK ARCHITECTURES}\label{NetA}
Following the discussion of section \ref{Dat}, we here explicit the normalization and log-scaling scheme. The clipped cut-out images (each of size $64 \times 64$) are log-scaled as:
$$x \longmapsto log_{10}(1+x) $$
x being the input image. Min-Max normalization is then performed for the log-scaled images :
\begin{equation}
    \text{Output} = \frac{\text{Input} - \text{min(Input)}}{\text{max(Input)}-\text{min(Input)}}
\end{equation}
where 'Output' is the normalized image, 'Input' is the original image, and 'min' and 'max' are functions that return the single minimal and maximal values of all the images in consideration. For generating images the min-max normalized image is further rescaled to the range [-1,1] as follows:
$$x \longmapsto (x-0.5)/0.5$$
It should be noted that the clipped cut-out images are directly Min-Max normalized before passing them to the classifier. It turns out they work better without log-scaling the images. 

Detailed information about the architecture of the implemented models is also provided in this appendix. The structure and corresponding number of parameters for the critic of the WGAN are shown in Table \ref{crtic_arch}, and for the generator in Table \ref{gen_arch}. Detailed information on CNN based classifiers can be found in Tables \ref{classifier_arch},\ref{classifier_arch_A} and for the Dense Network in Table \ref{classifier_archD}. Table \ref{tab:hyper} summarizes the hyperparameters for all model training conducted in this study. As the architecture for our DDPM is exactly the same as that of the original paper (with just final channel numbers being different), we refer the interested readers to the appendix B of \mbox{\citet{2020_Diff}} for relevant parameter details.
\setcounter{table}{0}
\renewcommand{\thetable}{A\arabic{table}}

% \begin{tabular}{c}
%     
%     Even sources labeled as cH or cR, which are considered uncertain, are included, as the classification notes that these sources still exhibit some central diffuse emission.\\ 
%     \hline
% \end{tabular}
%     
% \end{table}
\section{Generative Models' Comparison Results}\label{GM_Res}
Here we present the results for generative models' comparisons. The training datasets for WGANs and DDPMs consist of 13,440 and 4,288 images (each of size (64, 64, 1)), respectively, generated from GLEAM images of 10 sources after augmentation. The smaller dataset for DDPMs is due to their reliance on clean images, making Gaussian noise unsuitable for augmentation in this case. The testing dataset for this analysis consists of the 4,288 images from the DDPMs training dataset. For visual inspection, the WGAN-generated and DDPM-generated sample images are displayed in Fig. (\ref{Gen_app2}). This appendix supports the discussion in Section \ref{Res1}.

\begin{table*}
    \centering
    \caption{Parameters of the WGAN critic: This table details the architecture of the WGAN Critic, including Embedding, convolutional (Conv), and fully connected (Dense) layers. It includes kernel size, stride, input channels, depth, activation functions, and total parameters, indicating the model's complexity.}
    \begin{tabular}{cccccccc}
    \hline
    Layer   &  Name   &  Kernel size  & Stride  & Input channels  & Depth  & Activation &  Parameters\\
    \hline
    1   &  Embedding   &  -  & -  & 1  & 64  & - &  640 \\ 
    2   &  Dense 1   &  -  & -  & 64  & 512  & - &  33280 \\ 
    3   &  Conv1   &  8  & 2  & 1  & 32  & LeakyReLU &  2080 \\ 
    4   &  Conv2   &  1  & 2  & 32  & 64  & LeakyReLU &  2112 \\ 
    5   &  Conv3   &  8  & 2  & 64  & 128  & LeakyReLU &  524 416 \\ 
    6   &  Conv4   &  1  & 2  & 128  & 256  & LeakyReLU &  524 416 \\ 
    7   &  Dense 2   &  -  & -  & 4096  & 1024  & - &  4 195 328 \\
    8   &  Dense 3   &  -  & -  & 1536  & 1024  & - &  1 573 888 \\
    9   &  Dense 4   &  -  & -  & 1024  & 1  & Sigmoid &  1025 \\
    Total parameters: & & & & & & & 6 365 793 \\
    \hline
    
    \end{tabular}
    \label{crtic_arch}
\end{table*}
\vfill
\begin{table*}
    \centering
    \caption{Parameters of the WGAN generator: The WGAN Generator structure is outlined, with an Embedding layer, transposed convolutional (ConvT), and Dense layers. The table lists each layer’s configuration, highlighting the higher parameter count compared to the Critic, reflecting its capacity for data generation.}
    \begin{tabular}{cccccccc}
    \hline
    Layer   &  Name   &  Kernel size  & Stride  & Input channels  & Depth  & Activation 
 &  Parameters\\
    \hline
    1   &  Embedding   &  -  & -  & 1  & 64  & - & 640 \\ 
    2   &  Dense 1   &  -  & -  & 64  & 16  & - &  1040 \\ 
    3   &  Dense 2   &  -  & -  & 128  & 4096  & - &  528 384 \\
    4   &  ConvT1   &  8  & 2  & 257  & 256  & ReLU &  4 210 944 \\ 
    5   &  ConvT2   &  1  & 2  & 256  & 128  & ReLU &  32 896 \\ 
    6   &  ConvT3   &  8  & 2  & 128  & 64  & ReLU & 524 352 \\ 
    7   &  ConvT4   &  1  & 2  & 64  & 1  & ReLU &   65 \\ 
    8   &  Dense 3   &  -  & -  & 4096  & 4096  & Tanh &  16 781 312 \\
    Total parameters: & & & & & & & 22 079 633 \\
    \hline  
    \end{tabular}
    \label{gen_arch}
\end{table*}
\begin{table*}
    \centering
        \caption{Parameters of the Classifier I: This table shows the architecture of Classifier I, which consists of convolutional (Conv), max pooling (MaxPool), and Dense layers. The parameter count illustrates the model's simplicity relative to others.}
    \begin{tabular}{cccccccc}
    \hline
    Layer   &  Name   &  Kernel size  & Stride  & Input channels  & Depth  & Activation 
 &  Parameters\\
    \hline 
    1   &  Conv1   &  3  & 1  & 1  & 8  & ReLU &  80 \\ 
    2   &  MaxPool   &  2  & -  & -  & -  & - &  - \\ 
    3   &  Conv2   &  3  & 1  & 8  & 16  & ReLU &  1168 \\ 
    4   &  MaxPool   &  2  & -  & -  & -  & - & - \\ 
    5   & Dense 1   &  -  & -  & 3600  & 16  & ReLU &  57616 \\
    6  &  Dense 2   &  -  & -  & 16  & 2  & Softmax &  34 \\
    Total parameters: & & & & & & &  58 898 \\
    \hline
    
    \end{tabular}
    \label{classifier_arch}
\end{table*}
\begin{table*}
    \centering
        \caption{Parameters of the Classifier A: Classifier A is more complex, featuring convolutional layers, a multi-head attention mechanism, and layer normalization. The parameter count underscores the model's ability to handle more intricate tasks compared to Classifier I.}
    \begin{tabular}{cccccccc}
    \hline
    Layer   &  Name   &  Kernel size  & Stride  & Input channels  & Depth  & Activation 
 & Parameters\\
    \hline 
    1   &  Conv1   &  3  & 1  & 4  & 64  & ReLU &  2368 \\ 
    2   &  MaxPool   &  2  & -  & -  & -  & - &  - \\ 
    3   &  Conv2   &  3  & 1  & 64  & 128  & ReLU &  73856 \\ 
    4   &  MaxPool   &  2  & -  & -  & -  & - & - \\ 
    5   &  Conv3   &  3  & 1  & 128  & 192  & ReLU &  221376 \\
    6 & Multi-Head Functional & - & - & 192 & - & - & 117344 \\
    7 & Layer Normalization & - & - & 32 & - & - &  64 \\
    8   & Dense 1   &  -  & -  & 8192  & 64  & ReLU &  524352 \\
    9  &  Dense 2   &  -  & -  & 64  & 2  & Softmax &  130 \\
    Total parameters: & & & & & & &  939 490 \\
    \hline
    
    \end{tabular}
    \label{classifier_arch_A}
\end{table*}
\begin{table*}
    \centering
        \caption{Parameters of the Classifier D: Classifier D consists of Dense layers, similar to those in the WGAN tables. The total parameters indicate a balanced architecture designed for feature extraction and classification.}
    \begin{tabular}{cccccccc}
    \hline
    Layer   &  Name   & Input channels  & Depth  & Activation 
 & Parameters\\
    \hline 
    1   &  Dense 1    & 4096  & 64  & ReLU & 262 208 \\
    2   &  Dense 2   & 64  & 32  & ReLU & 2080 \\ 
    3   & Dense 3   & 32  & 16  & ReLU & 528 \\
    4  &  Dense 4   & 16  & 2  & Softmax & 34 \\
    Total parameters:  & & & & &  264 580 \\
    \hline
    \end{tabular}
    \label{classifier_archD}
\end{table*}
\begin{table*}
    \centering
    \caption{Hyperparameters of the training: Includes batch size, learning rate, optimizer (Adam), and $\beta_1,\beta_2$
  (which control gradient and variance decay). It also lists training time and iterations.}
    \begin{tabular}{cccccccc}
    \hline
         & Batch size & Learning rate & Optimizer & $\beta_1$ & $\beta_2$ & Training time & Iterations  \\
    \hline
    WGAN & 32 & 0.0002 & Adam & 0.5 & 0.999 &  $\approx$ 1-2 GPU hours & 10000 \\
    DDPM & 32 & 0.0002 with a scheduler & Adam & 0.9 & 0.999 &  $\approx$ 6.5 GPU hours & 46410\\
    Classifier {I,A,D} & 128 & 0.001 & Adam & 0.9 & 0.999 &  1-3 minutes on GPU & 1905\\
    \hline
    \end{tabular}
    \label{tab:hyper}
\end{table*}
\vspace{-8 pt}
\begin{figure*}
\centering
\hspace*{-0.5cm}
\begin{tabular}{cc}
\includegraphics[width=0.5\linewidth]{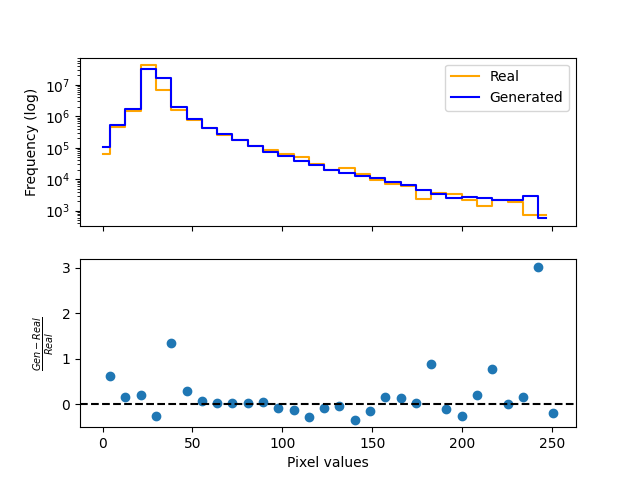}&
\includegraphics[width=0.5\linewidth]{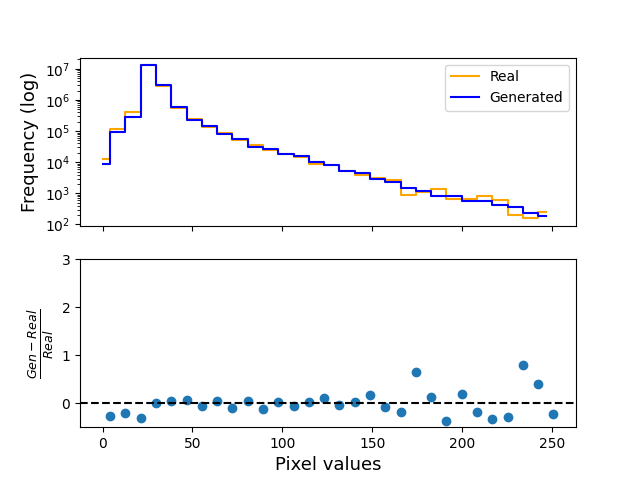}
\end{tabular}
\caption{Pixel value distributions for generated (blue) and real images (orange), with per-bin relative error in the bottom panels for WGANs (left) and DDPMs (right) trained on Single-Frequency Images}
\label{Gen_app1}
\end{figure*}

\begin{figure*}
\centering
\hspace*{-0.5cm}
\begin{tabular}{cc}
\includegraphics[width=0.5\linewidth]{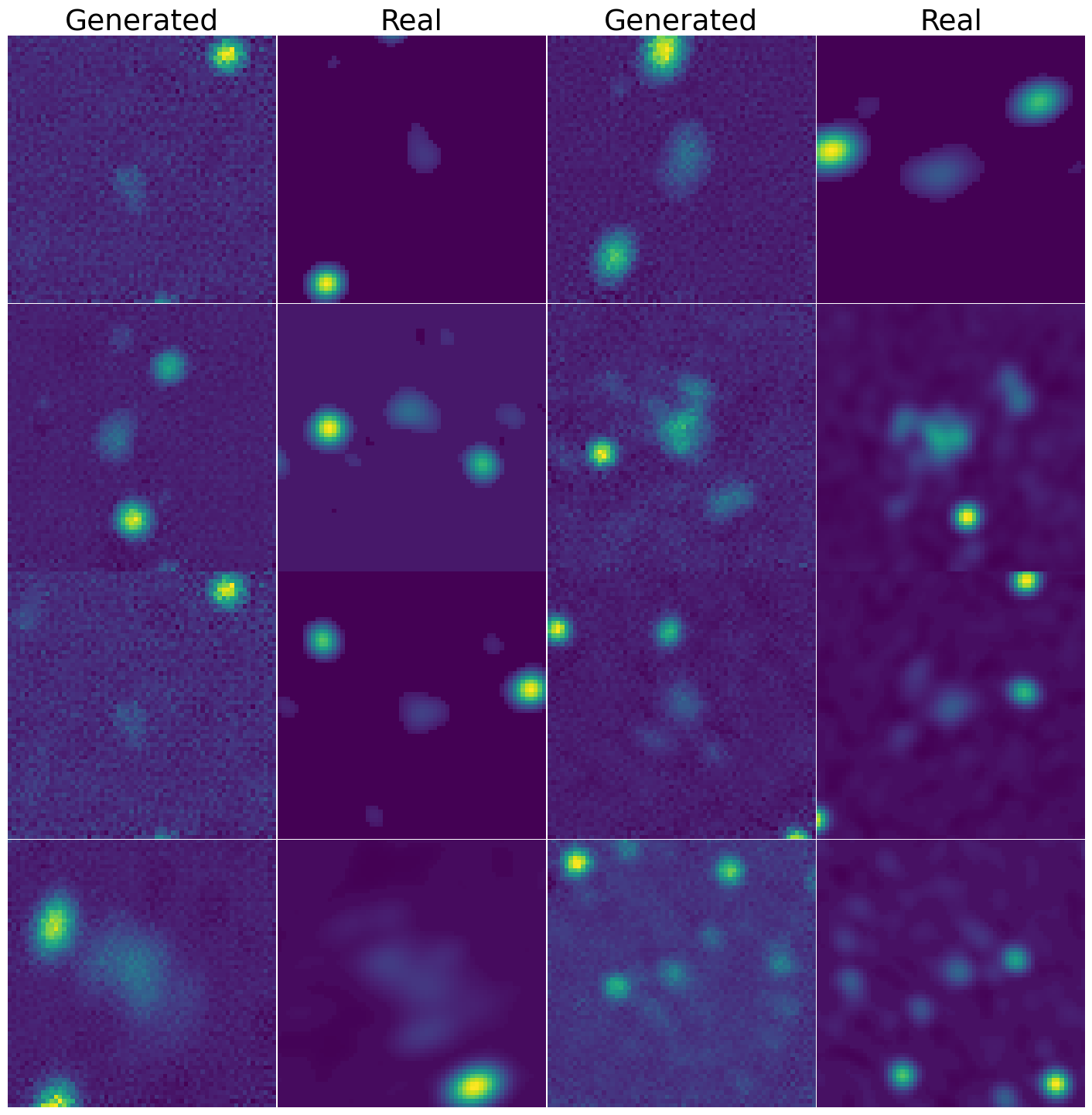}&
\includegraphics[width=0.5\linewidth]{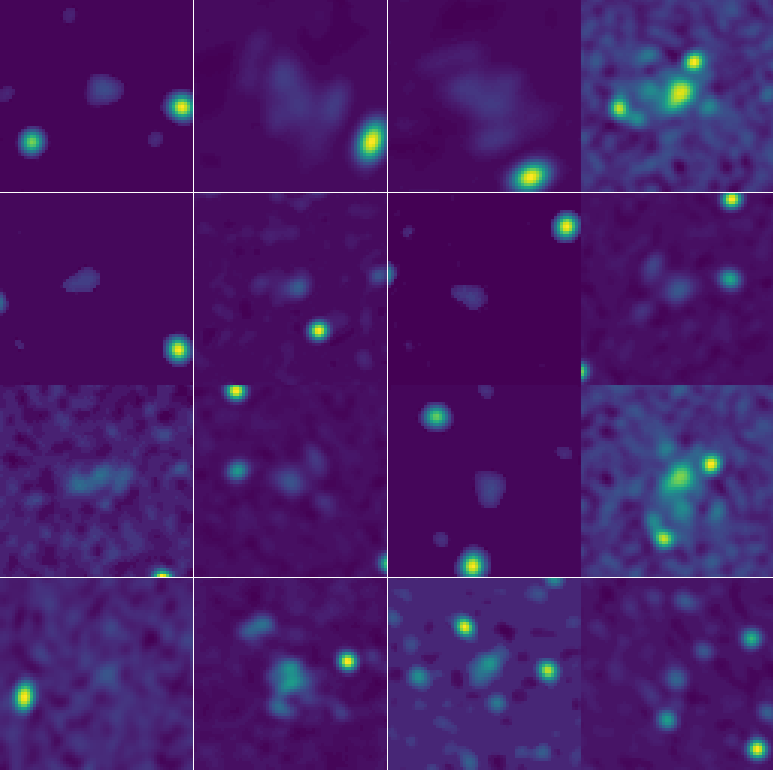}
\end{tabular}
\caption{Set of augmented Real and GAN Generated images side by side (left); Set of augmented unconditional diffusion-based generated images (right) }
\label{Gen_app2}

\end{figure*}

\end{appendices}

\bsp	% typesetting comment
\label{lastpage}
\end{document}